\documentclass[12pt]{article}
\usepackage{amsfonts,graphicx,amsmath,amsthm,amssymb,epsfig}
\usepackage{float}
\allowdisplaybreaks
\begin{document}

\title{\bf Study of Charged Compact Stars in Non-minimally Coupled Gravity}
\author{M. Sharif$^1$ \thanks{msharif.math@pu.edu.pk} and Tayyab Naseer$^{1,2}$ \thanks{tayyabnaseer48@yahoo.com}\\
$^1$ Department of Mathematics and Statistics, The University of Lahore,\\
1-KM Defence Road Lahore, Pakistan.\\
$^2$ Department of Mathematics, University of the Punjab,\\
Quaid-i-Azam Campus, Lahore-54590, Pakistan.}

\date{}
\maketitle

\begin{abstract}
This paper studies the structural formation of various spherically
symmetric anisotropic configured stars in
$f(\mathcal{R},\mathcal{T},\mathcal{Q})$ gravity under the influence
of electromagnetic field, where
$\mathcal{Q}=\mathcal{R}_{\eta\sigma}\mathcal{T}^{\eta\sigma}$. We
construct modified field equations by adopting the Krori-Barua
metric potentials (involving unknowns ($A,B,C$)) and employ bag
model equation of state to explore the physical characteristics of
compact structures like Vela X-I,~4U 1820-30,~SAX J 1808.4-3658,~Her
X-I and RXJ 1856-37. The bag constant ($\mathfrak{B_c}$) and other
three unknowns are calculated by using experimental data of all
considered stars. Further, we adopt a standard model
$\mathcal{R}+\varpi\mathcal{R}_{\eta\sigma}\mathcal{T}^{\eta\sigma}$
to interpret the graphical behavior of energy density, pressure in
radial and tangential directions as well as anisotropy. We also
investigate mass, compactness parameter, surface redshift and energy
conditions for compact bodies by taking model parameter ($\varpi$)
as $\pm5$. The stability of the obtained solution is examined by
means of two different criteria. We conclude that our proposed
solution meets all the physical requirements and shows stable
behavior everywhere for $\varpi=5$.
\end{abstract}

\section{Introduction}

General theory of Relativity (GR) has made significant breakthrough
in revealing numerous hidden components of the cosmos, yet it is not
sufficiently enough to examine our universe at large scale. Several
modifications to GR have recently been proposed to deal with the
puzzling issues related to cosmic evolution such as rapid expansion
and dark matter. This expanding phenomenon leads to the existence of
a mysterious kind of force which is immensely repulsive in nature,
called dark energy. This is why modified theories are labeled as
extremely important to unveil the cosmic' obscure features. The
$f(\mathcal{R})$ theory is the simplest modification to GR,
involving higher-order curvature terms owing to the replacement of
the Ricci scalar $\mathcal{R}$ with its generic function in the
Einstein-Hilbert action. Multiple approaches have been employed to
study the physical feasibility of various stellar configurations
\cite{2,3}. Capozziello \emph{et al.} \cite{8} investigated the
stability of some particular mathematical models by means of the
Lan\'{e}-Emden equation in this framework. Various researchers
\cite{9,9g} have done a comprehensive analysis in the study of
structural composition and evolution of astronomical bodies.

The lust to understand more fascinating features of our cosmos urged
Bertolami \emph{et al.} \cite{10} to study the effects of
matter-geometry coupling on massive bodies in $f(\mathcal{R})$
gravity. They proposed the idea of coupling by considering the
matter Lagrangian in terms of $\mathcal{R}$ and $\mathcal{L}_{m}$.
Such interaction between matter distribution and geometry of the
spacetime prompted many researchers to pay their attention on the
study of cosmic accelerating expansion. A number of modified
theories proposed in the last decade involving arbitrary coupling
have become a topic of great interest for astrophysicists. The first
theory which couples matter and geometry in some general manner at
the action level was proposed by Harko \emph{et al.} \cite{20},
named as $f(\mathcal{R},\mathcal{T})$ gravity in which $\mathcal{T}$
is trace of the energy-momentum tensor $(\mathrm{EMT})$. The
modified theories whose functional involve $\mathcal{T}$ result in
the non-conserved $\mathrm{EMT}$ which may lead to rapid cosmic
expansion. Numerous researchers \cite{21}-\cite{21f} studied
celestial objects in the framework of $f(\mathcal{R},\mathcal{T})$
theory and concluded that this theory provides many intriguing
results in astrophysics. This theory was further extended to a more
complex function $f(\mathcal{R},\mathcal{T},\mathcal{Q})$ by Haghani
\emph{et al.} \cite{22}, in which $\mathcal{Q}$ is contraction of
$\mathrm{EMT}$ and the Ricci tensor (i.e., $\mathcal{Q}\equiv
\mathcal{R}_{\eta\sigma}\mathcal{T}^{\eta\sigma}$) which helps to
entail non-minimal interaction even for the case when $\mathrm{EMT}$
is traceless. They examined physical feasibility of three different
mathematical models and also derived conserved equations of motion
through Lagrange multiplier method. Sharif and Zubair \cite{22a}
studied thermodynamical laws for black holes by assuming matter
Lagrangian as $L_m=\mu$ and $-P$. They also evaluated energy bounds
corresponding to some particular models in this framework
\cite{22b}.

Odintsov and S\'{a}ez-G\'{o}mez \cite{23} studied the occurrence of
$\Lambda$CDM model as well as several cosmological solutions in
$f(\mathcal{R},\mathcal{T},\mathcal{Q})$ scenario. It was also
stressed that the problem of matter instability is much complicated
in this theory due to the presence of higher-order terms in the
field equations. Ayuso \emph{et al.} \cite{24} adopted some
appropriate scalar as well as vector fields to find stability
conditions for massive structures and deduced that for the case of
vector field, the matter instability must occur. Baffou \emph{et
al.} \cite{25} solved modified field equations by means of
perturbation functions and found that these solutions are physically
viable. The isotropic and anisotropic configured neutron stars have
thoroughly been studied by Sharif and Waseem \cite{25a,25b}. Yousaf
\emph{et al.} \cite{26}-\cite{26c} decomposed the Riemann tensor
orthogonally and computed four structure scalars corresponding to
the charged/uncharged spherical geometry which are helpful in the
study of evolution of self-gravitating structures. This work has
also been extended for the case of cylindrical system
\cite{26d}-\cite{26eb}. Sharif and Naseer \cite{26f}-\cite{26j}
determined the solution of the charged/uncharged field equations
through multiple schemes and concluded that this theory provides
more stable results.

The relaxation of the stringent condition of isotropic fluid
distribution makes that system anisotropic, constituting a more
realistic situation according to the astrophysical viewpoint
\cite{26k}. This article considers anisotropic matter distribution,
therefore we may consider some other possible features in the inner
stellar composition, for example an electric charge. Its existence
in cosmological structures makes it easy to understand their
expansion and feasibility in a proper way. The first ever solution
to the charged interior was reported by Bonnor \cite{26ka}. Numerous
research has been carried out in the framework of GR and other
modified theories to investigate how charge affects physical
characteristics of celestial objects. Das \emph{et al.} \cite{27}
solved the Einstein-Maxwell equations after matching the interior
geometry with exterior Reissner-Nordstr\"{o}m at hypersurface. Sunzu
\emph{et al.} \cite{27a} utilized mass-radius relation to study
charge strange quark objects. Murad \cite{28} examined the effect of
charge on anisotropic compact stars and interpreted their physical
characteristics. Many authors \cite{28a}-\cite{28h} have done a
detailed study on various charged stellar systems and observed that
the presence of charge make these structures more stable.

The astronomical objects, particularly stars, play a crucial role in
the composition of galaxies in our cosmos. The structure of such
objects attracted numerous astrophysicists to concentrate on the
analysis of their evolutionary stages. The force of gravitation
exerted by a star due to its mass is always inward directed, while
the pressure produced by some nuclear reactions occurring in the
star's core is outward directed. They try to counterbalance the
gravitational pull and helps to hold the star in hydrostatic
equilibrium. Gravitational collapse occurs when pressure in outward
direction is no longer sufficient to counterbalance the
gravitational force, which results in the death of a stellar body
and the emergence of new astrophysical objects such as white dwarfs,
neutron stars and black holes according to the mass of dying star.
Many astronomers have studied the diverse characteristics of these
compact objects, among them neutron stars gained considerable
attention owing to their fascinating structural features and
evolution. The degeneracy pressure produced by neutrons
counterbalances the gravitational pull which results in the
hydrostatic equilibrium of these stars. Quark star is a highly dense
structure in between neutrons star and black hole, made up of
strange, up and down quark matter. These speculative structures and
their formation have been studied by several researchers
\cite{29a,29h}.

The study of physical properties of heavily bodies having
anisotropic configurations in their interiors becomes an influential
topic for numerous astrophysicists. The compact object which
contains density much higher than the nuclear density in its
interior should possess anisotropic matter distribution \cite{29i}.
The convincing effects of anisotropy on celestial structures have
been discussed by Herrera and Santos \cite{30}. Harko and Mak
\cite{31} studied static relativistic bodies and determined the
corresponding interior solutions by assuming particular form of
anisotropy. Hossein \emph{et al.} \cite{32} examined the influence
of anisotropy as well as cosmological constant $\Lambda$ on the
stability of massive self-gravitating systems. Kalam \emph{et al.}
\cite{32a} investigated physical feasibility of various neutron
stars by analyzing the validity of corresponding energy conditions
and stability criteria. Paul and Deb \cite{33} constructed some
physically feasible solutions for anisotropic configured compact
objects.

The equation of state (EoS) provided by MIT bag model is considered
as a useful tool to interrelate physical variables involved in the
interior matter distribution of quark bodies \cite{27,27a}. However,
the compactness of dense structures like SAX J 1808.4-3658, 4U
1820-30, RXJ 185635-3754, 4U 1728-34, PSR 0943+10 and Her X-1, etc.,
can successfully be explained by bag model which represents quark
matter EoS \cite{33a}, while EoS corresponding to the neutron star
fails in this case. The bag constant ($\mathfrak{B_c}$) helps to
determine the discrepancy between true and false states of a vacuum
and the increment in its value results in decreasing quark pressure.
The MIT bag model EoS has widely been used by several researchers
\cite{33b,34a} to examine the interior distribution of quark stars.
Rahaman \emph{et al.} \cite{35} employed an interpolating function
to determine the mass of various stars and also checked some
fundamental properties of a strange candidate with radius as 9.9km.
Bhar \cite{36} developed a hybrid star model by using Krori-Barua
metric potentials and found its calculated mass to be in agreement
with the observational one. Arba\~{n}il and Malheiro \cite{37}
studied the impact of anisotropy on various compact stars and
calculated numerical solution of corresponding radial perturbation,
hydrostatic equilibrium condition as well as MIT bag model. Deb
\emph{et al.} \cite{37a,37b} constructed some singularity-free
solutions to charged as well as uncharged strange bodies through
this EoS and interpreted them graphically in order to observe their
feasibility. Sharif and his colleagues \cite{38}-\cite{38f} extended
this work by means of MIT bag model and developed stable anisotropic
solutions in $f(\mathcal{R},\mathcal{T})$ and Brans-Dicke theories.

This paper examines various anisotropic quark star candidates and
their physical feasibility in view of the electromagnetic field by
considering a viable model of
$f(\mathcal{R},\mathcal{T},\mathcal{Q})$ theory. This paper is
organized as follows. Section 2 presents the fundamental formalism
of modified theory and the field equations involving MIT bag
constant. We assume Krori-Barua metric potentials involving three
unknowns which are calculated at the hypersurface in section 3. The
graphical interpretation of physical attributes for each star
candidate is given in section 4. Lastly, we sum up our results in
section 5.

\section{The $f(\mathcal{R},\mathcal{T},\mathcal{Q})$ Gravity}

The Einstein-Hilbert action becomes after inserting the complex
analytical functional $f$ in place of $\mathcal{R}$ (with
$\kappa=8\pi$) as \cite{23}
\begin{equation}\label{g1}
S_{f(\mathcal{R},\mathcal{T},\mathcal{R}_{\eta\sigma}\mathcal{T}^{\eta\sigma})}=\int
\left[\frac{f(\mathcal{R},\mathcal{T},\mathcal{R}_{\eta\sigma}\mathcal{T}^{\eta\sigma})}{16\pi}
+\mathcal{L}_{m}+\mathcal{L}_{\mathcal{E}}\right]\sqrt{-g}d^{4}x,
\end{equation}
where $\mathcal{L}_{\mathcal{E}}$ and $\mathcal{L}_{m}$ indicate the
Lagrangian densities of electromagnetic field and matter
configuration, respectively. The field equations agreeing with the
action \eqref{g1} take the form as
\begin{equation}\label{g2}
\mathcal{G}_{\eta\sigma}=8\pi\mathcal{T}_{\eta\sigma}^{(eff)}=-\frac{8\pi}{\mathcal{L}_{m}f_{\mathcal{Q}}
-f_{\mathcal{R}}}\big(\mathcal{T}_{\eta\sigma}+\mathcal{E}_{\eta\sigma}\big)+\mathcal{T}_{\eta\sigma}^{(\mathcal{D})}.
\end{equation}
The geometry of massive bodies is expressed by the term
$\mathcal{G}_{\eta\sigma}$ whereas
$\mathcal{T}_{\eta\sigma}^{(eff)}$ is pointed out as the
$\mathrm{EMT}$ in $f(\mathcal{R},\mathcal{T},\mathcal{Q})$ gravity
involving state variables and their derivatives in conjunction with
modified correction terms. The sector
$\mathcal{T}_{\eta\sigma}^{(\mathcal{D})}$ appears due to the
modified gravity is given as
\begin{eqnarray}\nonumber
\mathcal{T}_{\eta\sigma}^{(\mathcal{D})}&=&-\frac{1}{\mathcal{L}_{m}f_{\mathcal{Q}}-f_{\mathcal{R}}}
\left[\left(f_{\mathcal{T}}+\frac{1}{2}\mathcal{R}f_{\mathcal{Q}}\right)\mathcal{T}_{\eta\sigma}
+\left\{\frac{\mathcal{R}}{2}(\frac{f}{\mathcal{R}}-f_{\mathcal{R}})-\mathcal{L}_{m}f_{\mathcal{T}}\right.\right.\\\nonumber
&-&\left.\frac{1}{2}\nabla_{\chi}\nabla_{\xi}(f_{\mathcal{Q}}\mathcal{T}^{\chi\xi})\right\}g_{\eta\sigma}
-\frac{1}{2}\Box(f_{\mathcal{Q}}\mathcal{T}_{\eta\sigma})-(g_{\eta\sigma}\Box-
\nabla_{\eta}\nabla_{\sigma})f_{\mathcal{R}}\\\label{g4}
&-&2f_{\mathcal{Q}}\mathcal{R}_{\chi(\eta}\mathcal{T}_{\sigma)}^{\chi}+\nabla_{\chi}\nabla_{(\eta}[\mathcal{T}_{\sigma)}^{\chi}f_{\mathcal{Q}}]
+2(f_{\mathcal{Q}}\mathcal{R}^{\chi\xi}+\left.f_{\mathcal{T}}g^{\chi\xi})\frac{\partial^2\mathcal{L}_{m}}
{\partial g^{\eta\sigma}\partial g^{\chi\xi}}\right],
\end{eqnarray}
where the partial derivatives are defined as
$f_{\mathcal{R}}=\frac{\partial
f(\mathcal{R},\mathcal{T},\mathcal{Q})}{\partial
\mathcal{R}},~f_{\mathcal{T}}=\frac{\partial
f(\mathcal{R},\mathcal{T},\mathcal{Q})}{\partial \mathcal{T}}$ and
$f_{\mathcal{Q}}=\frac{\partial
f(\mathcal{R},\mathcal{T},\mathcal{Q})}{\partial \mathcal{Q}}$.
Also, the term $\nabla_\eta$ shows the mathematical notation of
covariant derivative and $\Box\equiv
g^{\eta\sigma}\nabla_\eta\nabla_\sigma$. As we considered the
Maxwell field, thus we take
$\mathcal{L}_{m}=-\frac{1}{4}\mathcal{F}_{\eta\sigma}\mathcal{F}^{\eta\sigma}$
which leads to $\frac{\partial^2\mathcal{L}_{m}} {\partial
g^{\eta\sigma}\partial
g^{\chi\xi}}=-\frac{1}{2}\mathcal{F}_{\eta\chi}\mathcal{F}_{\sigma\xi}$
\cite{22}. Here,
$\mathcal{F}_{\eta\sigma}=\omega_{\sigma;\eta}-\omega_{\eta;\sigma}$
is the Maxwell field tensor in which
$\omega_{\sigma}=\omega(r)\delta^{\sigma}_{0}$ represents the four
potential. In this theory, the divergence of $\mathrm{EMT}$ does not
vanish (i.e., $\nabla_\eta \mathcal{T}^{\eta\sigma}\neq 0$)
dissimilar to GR and $f(\mathcal{R})$ gravity as a result of the
existence of arbitrary matter-geometry coupling which violates the
equivalence principle. This results in the existence of an extra
force in spacetime geometry which hampers the test particles from
their geodesic motion in their gravitational field. Thus we have
\begin{align}\nonumber
\nabla^\eta
\big(\mathcal{T}_{\eta\sigma}+\mathcal{E}_{\eta\sigma}\big)&=\frac{2}{2f_\mathcal{T}+\mathcal{R}f_\mathcal{Q}+16\pi}
\bigg[\nabla_\eta(f_\mathcal{Q}\mathcal{R}^{\chi\eta}\mathcal{T}_{\chi\sigma})+\nabla_\sigma(\mathcal{L}_mf_\mathcal{T})\\\nonumber
&-\frac{1}{2}
(f_\mathcal{T}g_{\chi\xi}+f_\mathcal{Q}\mathcal{R}_{\chi\xi})\nabla_\sigma
\mathcal{T}^{\chi\xi}-\mathcal{G}_{\eta\sigma}\nabla^\eta(f_\mathcal{Q}\mathcal{L}_m)\\\label{g11}
&-\frac{1}{2}\big\{\nabla^{\eta}(\mathcal{R}f_{\mathcal{Q}})+2\nabla^{\eta}f_{\mathcal{T}}\big\}\mathcal{T}_{\eta\sigma}\bigg].
\end{align}

The interior fluid configuration of the astrophysical objects is
described by $\mathrm{EMT}$ whose each non-null ingredient discloses
particular physical attributes. The difference of tangential and
radial pressure components generates anisotropy in the inner region
of self-gravitating celestial structures which is noticed as the
crucial factor to examine their composition and expansion. A number
of heavily objects existing in our cosmos have been found to be
associated with anisotropic distribution, therefore this element has
persuasive consequences in the developmental phases of stellar
systems. We thus take anisotropic configuration which is represented
by the $\mathrm{EMT}$ defined as
\begin{equation}\label{g5}
\mathcal{T}_{\eta\sigma}=(\mu+P_\bot) \mathcal{K}_{\eta}
\mathcal{K}_{\sigma}+P_\bot
g_{\eta\sigma}+\left(P_r-P_\bot\right)\mathcal{W}_\eta\mathcal{W}_\sigma,
\end{equation}
where the quantities $\mathcal{K}_{\eta},~P_r,~P_\bot,$ and
$\mathcal{W}_\eta$ describe the four-velocity, radial pressure,
tangential pressure and the four-vector, respectively. The field
equations in this scenario provide the trace as follows
\begin{align}\nonumber
&3\nabla^{\chi}\nabla_{\chi}
f_\mathcal{R}-\mathcal{R}\left(\frac{\mathcal{T}}{2}f_\mathcal{Q}-f_\mathcal{R}\right)-\mathcal{T}(8\pi+f_\mathcal{T})+\frac{1}{2}
\nabla^{\chi}\nabla_{\chi}(f_\mathcal{Q}\mathcal{T})+\nabla_\eta\nabla_\chi(f_\mathcal{Q}\mathcal{T}^{\eta\chi})\\\nonumber
&-2f+(\mathcal{R}f_\mathcal{Q}+4f_\mathcal{T})\mathcal{L}_m+2\mathcal{R}_{\eta\chi}\mathcal{T}^{\eta\chi}f_\mathcal{Q}
-2g^{\sigma\xi} \frac{\partial^2\mathcal{L}_m}{\partial
g^{\sigma\xi}\partial
g^{\eta\chi}}\left(f_\mathcal{T}g^{\eta\chi}+f_\mathcal{Q}R^{\eta\chi}\right)=0.
\end{align}
The strong coupling between the geometry and matter of stellar
object can be disappeared when $\mathcal{Q}=0$ is taken in the
overhead equation, ultimately producing $f(\mathcal{R},\mathcal{T})$
theory, whereas the implementation of vacuum yields corresponding
results in $f(\mathcal{R})$ gravity. The electromagnetic
$\mathrm{EMT}$ has the form
\begin{equation*}
\mathcal{E}_{\eta\sigma}=\frac{1}{4\pi}\left[\frac{1}{4}g_{\eta\sigma}\mathcal{F}^{\rho\chi}\mathcal{F}_{\rho\chi}
-\mathcal{F}^{\chi}_{\eta}\mathcal{F}_{\sigma\chi}\right],
\end{equation*}
where $\mathcal{F}_{\rho\chi}$ must satisfy the Maxwell equations in
tensorial form written as
\begin{equation*}
\mathcal{F}^{\chi\eta}_{;\eta}=4\pi \mathcal{J}^{\chi}, \quad
\mathcal{F}_{[\chi\eta;\lambda]}=0,
\end{equation*}
$\mathcal{J}^{\chi}$ is the current density which can further be
expressed in terms of charge density $\zeta$ as
$\mathcal{J}^{\chi}=\zeta \mathcal{K}^{\chi}$.

The hypersurface $\Sigma$ separates the interior and exterior
regions of self-gravitating geometry. The following metric expresses
static spherically symmetric matter configuration corresponding to
the interior geometry as
\begin{equation}\label{g6}
ds^2=-e^{\phi} dt^2+e^{\varphi} dr^2+r^2d\theta^2+r^2\sin^2\theta
d\psi^2,
\end{equation}
where $\phi=\phi(r)$ and $\varphi=\varphi(r)$. The Maxwell field
equations corresponding to the above metric become
\begin{equation}
\omega''+\frac{1}{2r}\big[4-r(\phi'+\varphi')\big]\omega'=4\pi\zeta
e^{\frac{\phi}{2}+\varphi},
\end{equation}
whose integration produces
\begin{equation}
\omega'=\frac{q}{r^2}e^{\frac{\phi+\varphi}{2}},
\end{equation}
where $q=q(r)$ represents charge in the interior geometry
\eqref{g6}. Here, $'=\frac{\partial}{\partial r}$. The matter
Lagrangian in this case becomes as
$\mathcal{L}_{m}=\frac{q^2}{2r^4}$. To continue our analysis, we
consider the comoving framework, thus four-velocity and four-vector
take the form as
\begin{equation}\label{g7}
\mathcal{K}^\eta=\delta^\eta_0 e^{\frac{-\phi}{2}}, \quad
\mathcal{W}^\eta=\delta^\eta_1 e^{\frac{-\varphi}{2}},
\end{equation}
which satisfy $\mathcal{K}^\eta \mathcal{K}_{\eta}=-1$ and
$\mathcal{W}^\eta \mathcal{K}_{\eta}=0$. The current phase of cosmic
accelerating expansion contains numerous stars that exist in
non-linear regime. We may better understand the formation of massive
structures by studying their linear behavior. The functional form of
$f(\mathcal{R},\mathcal{T},\mathcal{Q})$ gravity is much complicated
as compared to other theories. We thus consider its separable form
proposed by Haghani \emph{et al.} \cite{22} to examine the impact of
$\mathcal{Q}=\mathcal{R}_{\eta\sigma}\mathcal{T}^{\eta\sigma}$ on
different strange star candidates as
\begin{equation}\label{g61}
f(\mathcal{R},\mathcal{T},\mathcal{R}_{\eta\sigma}\mathcal{T}^{\eta\sigma})=f_1(\mathcal{R})+
f_2(\mathcal{R}_{\eta\sigma}\mathcal{T}^{\eta\sigma})=\mathcal{R}+\varpi\mathcal{R}_{\eta\sigma}\mathcal{T}^{\eta\sigma},
\end{equation}
where $\varpi$ represents an arbitrary constant.

In this case, the positive values of $\varpi$ offer the solution
that possesses an oscillatory (collapsing and expanding) behavior.
On the other hand, the cosmic scale factor has the hyperbolic
cosine-type dependence when $\varpi<0$. Numerous studies on the
stability and viability of multiple isotropic and anisotropic
solutions have been done by utilizing this model
\cite{22a,22b,25a,25b}. Since, $\varpi$ is a real-valued constant
that couples the matter and geometric terms, one can take any value
(either positive or negative) to check the physical acceptability of
the solution to the field equations. In view of this, we consider
$\varpi=\pm5$ throughout the analysis. Here,
\begin{eqnarray}\nonumber
\mathcal{Q}&=&e^{-\varphi}\bigg[\frac{\mu}{4}\left(2\phi''-\phi'\varphi'+\phi'^2+\frac{4\phi'}{r}\right)
-\frac{P_r}{4}\left(2\phi''-\phi'\varphi'+\phi'^2+\frac{4\varphi'}{r}\right)\\\nonumber
&+&P_\bot\left(\frac{2e^\varphi}{r^2}-\frac{\phi'}{r}+\frac{\varphi'}{r}-\frac{2}{r^2}\right)\bigg].
\end{eqnarray}
Combining Eqs.\eqref{g2} and \eqref{g4}, we obtain the following
expression in terms of the particular model \eqref{g61} as
\begin{eqnarray}\nonumber
\mathcal{G}_{\eta\sigma}&=&\frac{1}{1-\frac{\varpi q^2}{2r^4}}
\bigg[\left(8\pi+\frac{1}{2}\varpi\mathcal{R}\right)\mathcal{T}_{\eta\sigma}+8\pi\mathcal{E}_{\eta\sigma}
+\frac{\varpi}{2}\left\{\mathcal{Q}-\nabla_{\chi}\nabla_{\xi}\mathcal{T}^{\chi\xi}\right\}g_{\eta\sigma}\\\label{g7a}
&-&\frac{\varpi}{2}\Box\mathcal{T}_{\eta\sigma}-2\varpi\mathcal{R}_{\chi(\eta}\mathcal{T}_{\sigma)}^{\chi}
+\varpi\nabla_{\chi}\nabla_{(\eta}\mathcal{T}_{\sigma)}^{\chi}+2\varpi\mathcal{R}^{\chi\xi}\frac{\partial^2\mathcal{L}_{m}}
{\partial g^{\eta\sigma}\partial g^{\chi\xi}}\bigg].
\end{eqnarray}
The non-conservation of $\mathrm{EMT}$ \eqref{g11} becomes after
inserting the considered model as
\begin{eqnarray}\nonumber
\nabla^\eta
\big(\mathcal{T}_{\eta\sigma}+\mathcal{E}_{\eta\sigma}\big)&=&\frac{2\varpi}{\varpi\mathcal{R}+16\pi}
\left[\nabla_\eta(\mathcal{R}^{\chi\eta}\mathcal{T}_{\chi\sigma})-\frac{1}{2}\mathcal{R}_{\chi\xi}\nabla_\sigma\mathcal{T}^{\chi\xi}-\frac{1}{2}
\mathcal{T}_{\eta\sigma}\nabla^\eta\mathcal{R}\right.\\\label{g7b}
&-&\left.\mathcal{G}_{\eta\sigma}\nabla^\eta\left(\frac{q^2}{2r^4}\right)\right].
\end{eqnarray}
The field equations \eqref{g7a} corresponding to the matter
distribution \eqref{g5} and geometry \eqref{g6} become
\begin{align}\nonumber
8\pi\mu&=e^{-\varphi}\bigg[\frac{\varphi'}{r}+\frac{e^\varphi}{r^2}-\frac{1}{r^2}+\varpi\bigg\{\mu\bigg(\frac{3\phi'\varphi'}{8}-\frac{\phi'^2}{8}
+\frac{\varphi'}{r}+\frac{e^\varphi}{r^2}-\frac{1}{r^2}-\frac{3\phi''}{4}\\\nonumber
&-\frac{3\phi'}{2r}\bigg)-\mu'\bigg(\frac{\varphi'}{4}-\frac{1}{r}-\phi'\bigg)+\frac{\mu''}{2}+P_r\bigg(\frac{\phi'\varphi'}{8}
-\frac{\phi'^2}{8}-\frac{\phi''}{4}+\frac{\varphi'}{2r}+\frac{\varphi''}{2}\\\nonumber
&-\frac{3\varphi'^2}{4}\bigg)+\frac{5\varphi'P'_r}{4}-\frac{P''_r}{2}+P_\bot\bigg(\frac{\varphi'}{2r}-\frac{\phi'}{2r}+\frac{e^\varphi}{r^2}
+\frac{1}{r^2}\bigg)-\frac{P'_\bot}{r}+\frac{q^2}{r^4}\bigg(\frac{\varphi'}{2r}\\\label{g8}
&-\frac{e^\varphi}{2r^2}+\frac{1}{2r^2}+\frac{\phi'\varphi'}{8}
-\frac{\phi'^2}{8}-\frac{\phi''}{4}-\frac{e^\varphi}{\varpi}\bigg)\bigg\}\bigg],\\\nonumber
8\pi
P_r&=e^{-\varphi}\bigg[\frac{\phi'}{r}-\frac{e^\varphi}{r^2}+\frac{1}{r^2}+\varpi\bigg\{\mu\bigg(\frac{\phi'\varphi'}{8}+\frac{\phi'^2}{8}
-\frac{\phi''}{4}-\frac{\phi'}{2r}\bigg)-\frac{\phi'\mu'}{4}-P_r\\\nonumber
&\times\bigg(\frac{5\phi'^2}{8}-\frac{7\phi'\varphi'}{8}+\frac{5\phi''}{4}-\frac{7\varphi'}{2r}+\frac{\phi'}{r}-\varphi'^2
-\frac{e^\varphi}{r^2}+\frac{1}{r^2}\bigg)+P'_r\bigg(\frac{\phi'}{4}+\frac{1}{r}\bigg)\\\nonumber
&-P_\bot\bigg(\frac{\varphi'}{2r}-\frac{\phi'}{2r}+\frac{e^\varphi}{r^2}
+\frac{1}{r^2}\bigg)+\frac{P'_\bot}{r}+\frac{q^2}{r^4}\bigg(\frac{\phi'}{2r}+\frac{e^\varphi}{2r^2}
-\frac{1}{2r^2}+\frac{\phi''}{4}+\frac{\phi'^2}{8}\\\label{g9}
&-\frac{\phi'\varphi'}{8}+\frac{e^\varphi}{\varpi}\bigg)\bigg\}\bigg],\\\nonumber
8\pi
P_\bot&=e^{-\varphi}\bigg[\frac{\phi'^2}{4}-\frac{\phi'\varphi'}{4}+\frac{\phi''}{2}-\frac{\varphi'}{2r}+\frac{\phi'}{2r}
+\varpi\bigg\{\mu\bigg(\frac{\phi'^2}{8}+\frac{\phi'\varphi'}{8}-\frac{\phi''}{4}-\frac{\phi'}{2r}\bigg)\\\nonumber
&-\frac{\phi'\mu'}{4}+P_r\bigg(\frac{\phi'^2}{8}-\frac{\phi'\varphi'}{8}+\frac{\phi''}{4}-\frac{\varphi'}{2r}-\frac{\varphi''}{2}
+\frac{3\varphi'^2}{4}\bigg)-\frac{5\varphi'P'_r}{4}+\frac{P''_r}{2}\\\nonumber
&-P_\bot\bigg(\frac{\phi'^2}{4}-\frac{\phi'\varphi'}{4}+\frac{\phi''}{2}-\frac{\varphi'}{r}+\frac{\phi'}{r}-\frac{2e^\varphi}{r^2}
+\frac{1}{r^2}\bigg)-P'_\bot\bigg(\frac{\varphi'}{4}-\frac{\phi'}{4}-\frac{3}{r}\bigg)\\\label{g10}
&+\frac{P''_\bot}{2}+\frac{q^2}{r^4}\bigg(\frac{\phi'\varphi'}{8}-\frac{\phi'^2}{8}-\frac{\phi''}{4}+\frac{\varphi'}{4r}
-\frac{\phi'}{4r}-\frac{e^\varphi}{\varpi}\bigg)\bigg\}\bigg].
\end{align}
The system of above equations becomes more complex as the state
variables and their derivatives appear. From Eq.\eqref{g7b}, we
obtain the hydrostatic equilibrium condition in modified scenario in
the presence of charge as
\begin{align}\nonumber
&\frac{dP_r}{dr}+\frac{\phi'}{2}\left(\mu
+P_r\right)-\frac{2}{r}\left(P_\bot-P_r\right)-\frac{qq'}{4\pi
r^4}-\frac{2\varpi
e^{-\varphi}}{\varpi\mathcal{R}+16\pi}\bigg[\frac{\phi'\mu}{8}\bigg(\phi'^2-\phi'\varphi'+2\phi''\\\nonumber
&+\frac{4\phi'}{r}\bigg)-\frac{\mu'}{8}\bigg(2\phi''-\phi'\varphi'+\phi'^2+\frac{4\phi'}{r}\bigg)+P_r\bigg(\frac{5\phi'^2\varphi'}{8}
-\frac{5\phi'\varphi'^2}{8}-\frac{5\varphi'^2}{2r}-\frac{\phi'''}{2}\\\nonumber
&-\phi'\phi''+\frac{7\phi''\varphi'}{4}+\frac{\phi'\varphi''}{2}+\frac{2\varphi''}{r}+\frac{\phi'\varphi'}{r}-\frac{\varphi'}{r^2}
-\frac{\phi''}{r}+\frac{\phi'}{r^2}+\frac{2e^\varphi}{r^3}-\frac{2}{r^3}\bigg)\\\nonumber
&+\frac{P'_r}{8}\bigg(\phi'\varphi'-2\phi''-\phi'^2+\frac{4\varphi'}{r}\bigg)+\frac{P_\bot}{r^2}\bigg(\varphi'-\phi'+\frac{2e^\varphi}{r}
-\frac{2}{r}\bigg)-\frac{P'_\bot}{r}\bigg(\frac{\varphi'}{2}-\frac{\phi'}{2}\\\label{g12}
&+\frac{e^\varphi}{r}-\frac{1}{r}\bigg)-\bigg(\frac{qq'}{r^4}-\frac{2q^2}{r^5}\bigg)\left(\frac{\phi'}{r}-\frac{e^\varphi}{r^2}+\frac{1}{r^2}
+\frac{2e^\varphi}{\varpi}\right)\bigg]=0.
\end{align}
This equation gives the most general form of
Tolman-Opphenheimer-Volkoff ($\mathrm{TOV}$) equation which seems to
be considerably significant in studying the dynamical properties of
celestial objects.

Several constraints, that help to interlink the matter variables of
the fluid configuration, termed as EoSs, are significant to
investigate the physical characteristics of a massive body. The
heavily bodies with masses 8 to 20 times mass of the sun turn into
neutron stars after collapse that are considered being the most
intriguing objects in this cosmos. They can further be transformed
into two different objects (such as quark stars or black holes)
according to their respective densities \cite{33b}. It is
interesting to note that these stars occupy strong gravitational
field due to highly dense nature but small in size. The under
determined system of the field equations \eqref{g8}-\eqref{g10} is
highly non-linear that contains six unknowns
$(\phi,\varphi,\mu,P_r,P_\bot,q)$, thus we require some constraints
to close the system. We assume the known form of the charge
suggested by de Felice \emph{et al.} \cite{39} to reduce the number
of unknowns. Further, we consider MIT bag model EoS which links the
state variables in the interior of compact geometry and plays
significant role in analyzing the properties of quark stars
\cite{27,27a}.

Although a fairly common supposition in the analysis of compact
structures is the isotropic fluid configuration, we find some strong
evidences which suggest the appearance of local anisotropy in these
objects \cite{40ba,40bg}. There are many possible sources in this
regard, one of them is associated to the strong magnetic fields
observed in celestial objects such as white dwarfs, neutron stars or
strange quark stars \cite{40c,40cd}. In this regard, the quark
pressure with a bag constant ($\mathfrak{B_c}$) is defined as
\begin{equation}\label{g13}
P_r+\mathfrak{B_c}=\sum_{l=u,d,s}P^l,
\end{equation}
where the terms $P^s,~P^d$ and $P^u$ correspond to the pressures of
strange, down and up quark fluids, respectively. The relation
$\mu^l=3P^l$ gives the relation between each quark pressure with its
respective quark density. Hence the energy density has the form
\begin{equation}\label{g14}
\mu-\mathfrak{B_c}=\sum_{l=u,d,s}\mu^l.
\end{equation}
Using Eqs.\eqref{g13} and \eqref{g14}, the MIT bag model EoS
corresponding to the strange fluid becomes
\begin{equation}\label{g14a}
P_r=\frac{1}{3}\left(\mu-4\mathfrak{B_c}\right).
\end{equation}
The analysis on physical features of anisotropic strange stars has
successfully been done for different calculated values of bag
constant \cite{41g}. Further, the analytic solution to the field
equations \eqref{g8}-\eqref{g10} is obtained by employing EoS
\eqref{g14a} as
\begin{align}\nonumber
\mu&=\bigg[8\pi
e^{\varphi}+\varpi\bigg(\frac{9\phi''}{8}-\frac{e^{\varphi}}{r^2}+\frac{1}{r^2}-\frac{\varphi''}{8}-\frac{5\phi'\varphi'}{8}-\frac{\varphi'^2}{16}
-\frac{7\varphi'}{2r}+\frac{3\phi'^2}{16}+\frac{7\phi'}{4r}\bigg)\bigg]^{-1}\\\nonumber
&\times\bigg[\frac{3}{4}\bigg(1+\frac{\varpi
q^2}{2r^4}\bigg)\bigg(\frac{\varphi'}{r}+\frac{\phi'}{r}\bigg)+\mathfrak{B_c}\bigg\{8\pi
e^\varphi-\varpi\bigg(\frac{4\varphi'}{r}-\frac{3\phi'^2}{4}-\frac{3\phi''}{2}+\phi'\varphi'\\\label{g14b}
&\frac{\varphi''}{2}+\frac{\varphi'^2}{4}-\frac{\phi'}{r}+\frac{e^\varphi}{r^2}-\frac{1}{r^2}\bigg)\bigg\}\bigg],\\\nonumber
P_r&=\bigg[8\pi
e^{\varphi}+\varpi\bigg(\frac{9\phi''}{8}-\frac{e^{\varphi}}{r^2}+\frac{1}{r^2}-\frac{\varphi''}{8}-\frac{5\phi'\varphi'}{8}-\frac{\varphi'^2}{16}
-\frac{7\varphi'}{2r}+\frac{3\phi'^2}{16}+\frac{7\phi'}{4r}\bigg)\bigg]^{-1}\\\nonumber
&\times\bigg[\frac{1}{4}\bigg(1+\frac{\varpi
q^2}{2r^4}\bigg)\bigg(\frac{\varphi'}{r}+\frac{\phi'}{r}\bigg)-\mathfrak{B_c}\bigg\{8\pi
e^\varphi-\varpi\bigg(\frac{\phi'\varphi'}{2}
+\frac{\varphi'}{r}-\frac{2\phi'}{r}+\frac{e^\varphi}{r^2}\\\label{g14c}
&-\phi''-\frac{1}{r^2}\bigg)\bigg\}\bigg],\\\nonumber
P_\bot&=\bigg[8\pi
e^{\varphi}+\varpi\bigg(\frac{\phi'^2}{4}-\frac{2e^{\varphi}}{r^2}+\frac{1}{r^2}-\frac{\phi'\varphi'}{4}+\frac{\phi''}{2}-\frac{\varphi'}{r}
+\frac{\phi'}{r}\bigg)\bigg]^{-1}\bigg[\frac{\phi'^2}{4}-\frac{\varphi'}{2r}\\\nonumber
&+\frac{\phi'}{2r}-\frac{\phi'\varphi'}{4}+\frac{\phi''}{2}+\varpi\bigg\{8\pi
e^{\varphi}+\varpi\bigg(\frac{9\phi''}{8}-\frac{e^{\varphi}}{r^2}+\frac{1}{r^2}-\frac{\varphi''}{8}
-\frac{5\phi'\varphi'}{8}-\frac{\varphi'^2}{16}\\\nonumber
&-\frac{7\varphi'}{2r}+\frac{3\phi'^2}{16}+\frac{7\phi'}{4r}\bigg)\bigg\}^{-1}\bigg\{\frac{1}{8r}\bigg(1+\frac{\varpi
q^2}{2r^4}\bigg)\bigg(2\phi'\varphi'^2+\phi'^3-\phi''\varphi'-\phi'\phi''\\\nonumber
&-\varphi'\varphi''-\phi'\varphi''+\frac{3\phi'^2\varphi'}{2}-\frac{3\phi'^2}{r}+\frac{3\varphi'^3}{2}
-\frac{\varphi'^2}{r}-\frac{4\phi'\varphi'}{r}\bigg)+2\pi
e^\varphi\mathfrak{B_c}\bigg(\phi'\varphi'\\\nonumber
&-2\phi''+2\varphi''-3\varphi'^2-\frac{2\phi'}{r}+\frac{2\varphi'}{r}\bigg)
+\frac{\varpi\mathfrak{B_c}}{16}\bigg(10\phi''\varphi''-5\phi'\varphi'\varphi''+11\phi'\phi''\varphi'\\\nonumber
&-11\phi''\varphi'^2-\phi'^2\varphi''
-2\phi''\phi'^2-10\phi''^2-\frac{7\phi'^2\varphi'^2}{2}+\frac{\phi'^3\varphi'}{2}-\frac{36\phi'\varphi'^2}{r}-\frac{8\phi'^3}{r}\\\nonumber
&+\frac{11\phi'\varphi'^3}{2}+\frac{16\phi'^2\varphi'}{r}+\frac{28\phi''\varphi'}{r}-\frac{8\varphi'\varphi''}{r}+\frac{12\varphi'^3}{r}
+\frac{3\phi'^4}{2}-\frac{8\phi'^2}{r^2}-\frac{8\varphi''e^\varphi}{r^2}\\\nonumber
&+\frac{8\varphi''}{r^2}-\frac{20\varphi'^2}{r^2}-\frac{24\phi'\phi''}{r}+\frac{52\phi'\varphi'}{r^2}+\frac{10\phi'\varphi''}{r}
-\frac{4e^\varphi\phi'\varphi'}{r^2}+\frac{8e^\varphi\phi''}{r^2}-\frac{8\phi''}{r^2}\\\nonumber
&+\frac{12\varphi'^2e^\varphi}{r^2}-\frac{8\phi'}{r^3}-\frac{8e^\varphi\varphi'}{r^3}+\frac{8\varphi'}{r^3}
+\frac{8e^\varphi\phi'}{r^3}\bigg)\bigg\}\bigg]+\frac{\varpi
q^2}{4r^4e^\varphi}\bigg(\frac{\phi'\varphi'}{2}-\frac{\phi'^2}{2}-\phi''\\\label{g14d}
&+\frac{\varphi'}{r}-\frac{\phi'}{r}-\frac{4e^\varphi}{\varpi}\bigg).
\end{align}
It is seen that the EoS \eqref{g14a} has widely been utilized in GR
as well as modified scenario to study the physical features of
various quark bodies.

Now, we have six unknowns
$(\phi,\varphi,\mu,P_r,P_\bot,\mathfrak{B_c})$ along with field
equations \eqref{g14b}-\eqref{g14d}. Our objective is to construct
anisotropic charged solution and figure out its feasibility for two
different values of the coupling constant $\varpi$ corresponding to
five star candidates. To accomplish this, we need to impose three
constraints, i.e., the vanishing radial pressure condition at the
boundary (that will be discussed in the next section) and a known
form of the metric potentials. Thus, we adopt metric potentials
proposed by Krori-Barua \cite{41h} in
$f(\mathcal{R},\mathcal{T},\mathcal{Q})$ scenario which was
initially used to study charged stellar structures and its
non-singular nature gained considerable interest in astrophysics.
The solution has the form
\begin{equation}\label{g15}
\phi=Br^2+C, \quad \varphi=Ar^2,
\end{equation}
which involves $A,~B$ and $C$ as unknown constants whose values can
be computed through boundary conditions. The field equations
\eqref{g14b}-\eqref{g14d} in terms of Krori-Barua ansatz \eqref{g15}
become
\begin{align}\nonumber
\mu&=\bigg[r^2\big\{\varpi\big(-A^2r^4-5Ar^2\big(2Br^2+3\big)+3B^2r^4+23Br^2+4\big)-4e^{Ar^2}\big(\varpi\\\nonumber
&-8\pi
r^2\big)\big\}\bigg]^{-1}\bigg[-4\varpi\mathfrak{B_c}r^2\big(A^2r^4+Ar^2\big(4Br^2+9\big)+e^{Ar^2}-3B^2r^4-5Br^2\\\label{g16}
&-1\big)+3(A+B)\big(\varpi
q^2+2r^4\big)+32\pi\mathfrak{B_c}r^4e^{Ar^2}\bigg],\\\nonumber
P_r&=\bigg[r^2\big\{\varpi\big(-A^2r^4-5Ar^2\big(2Br^2+3\big)+3B^2r^4+23Br^2+4\big)-4e^{Ar^2}\big(\varpi\\\nonumber
&-8\pi
r^2\big)\big\}\bigg]^{-1}\bigg[A\big(8\varpi\mathfrak{B_c}Br^6+\varpi
q^2+r^4(8\varpi\mathfrak{B_c}
+2)\big)-4\mathfrak{B_c}r^2\big(\varpi+e^{Ar^2}\\\label{g17}
&\times\big(8\pi r^2-\varpi \big)\big)+B\big(\varpi
q^2+r^4(2-24\varpi\mathfrak{B_c})\big) \bigg],\\\nonumber
P_\bot&=\frac{1}{4}e^{-Ar^2}\bigg[4\big(-A(1+Br^2)+B(2+Br^2)\big)\bigg\{8\pi+\varpi
e^{-Ar^2}\bigg(3B+\frac{1-e^{Ar^2}}{r^2}\\\nonumber
&+B^2r^2-A(2+Br^2)\bigg)\bigg\}^{-1}+\frac{2q^2}{r^4}\big(\varpi\big(A+ABr^2-B\big(2+Br^2\big)\big)-2e^{Ar^2}\big)\\\nonumber
&+16r^4\varpi e^{Ar^2}\bigg\{\frac{1}{4r^2}\big((A+B)
\big(3A^2r^2+A\big(Br^2-2\big)+2B\big(Br^2-2\big)\big)\big(\varpi
q^2\\\nonumber
&+2r^4\big)\big)+8\pi\mathfrak{B_c}e^{Ar^2}\big(A\big(2+Br^2\big)-2B-3A^2r^2\big)
+\frac{\varpi\mathfrak{B_c}}{2r^2}\big(A^3r^4\big(11Br^2\\\nonumber
&+12\big)-A^2r^2\big(-6e^{Ar^2}+7B^2r^4+52Br^2+14\big)+B\big(4e^{Ar^2}+3B^3r^6-4\\\nonumber
&-10B^2r^4-21Br^2\big)+A\big(-2e^{Ar^2}\big(Br^2+2\big)+B^3r^6+18B^2r^4+55Br^2\\\nonumber
&+4\big)\big)\bigg\}\bigg\{\big(\varpi\big(-Ar^2\big(Br^2+2\big)+B^2r^4+3Br^2+1\big)+e^{Ar^2}\big(8\pi
r^2-\varpi\big)\big)\\\nonumber
&\times\big(\varpi\big(-A^2r^4-5Ar^2\big(2Br^2+3\big)+3B^2r^4+23Br^2+4\big)+4e^{Ar^2}\big(8\pi
r^2\\\label{g18} &-\varpi\big)\big) \bigg\}^{-1}\bigg].
\end{align}

\section{Boundary Conditions}

Junction conditions play a vital role in understanding the physical
characteristics of the celestial objects at the boundary $(\Sigma)$.
The fundamental forms of Darmois junction conditions are given as
follows \cite{41ja}
\begin{itemize}
\item The first form states that the continuity of the metric
potentials of both inner and outer geometries holds at the $\Sigma$.
\item The second form points out that there is no difference between the extrinsic curvature of
both spacetimes that yields $P_r{_=^\Sigma}0$.
\end{itemize}
One should choose the outer geometry on the basis that the
fundamental properties (such as charged/uncharged and
static/non-static, etc.) of the outer and inner spacetimes can match
with each other at the spherical boundary. Since the inner geometry
\eqref{g6} is influenced by an electromagnetic field, thus the most
suitable choice for an outer spacetime in this regard is the
Reissner-Nordstr\"{o}m metric. The junction conditions in
$f(\mathcal{R})$ theory are different from that of GR due to the
involvement of higher-order geometric terms \cite{41jaa,41jaaa}. For
instance, the Starobinsky model contains the Ricci scalar up to the
second order given by
$f(\mathcal{R})=\mathcal{R}+\alpha\mathcal{R}^2$, where $\alpha$ is
a restricted model parameter. However, for the model \eqref{g61},
the inclusion of curvature term $\mathcal{R}$ represents GR and
there is no contribution of the term
$\mathcal{R}_{\eta\sigma}\mathcal{T}^{\eta\sigma}$ to the current
scenario. Thus, the exterior geometry in this case will be that of
GR. The outer geometry is given by
\begin{equation}\label{g20}
ds^2=-\left(1-\frac{2\bar{M}}{r}+\frac{\bar{Q}^2}{r^2}\right)dt^2+\frac{dr^2}{\left(1-\frac{2\bar{M}}{r}+\frac{\bar{Q}^2}{r^2}\right)^{-1}}
+r^2d\theta^2+r^2\sin^2\theta d\psi^2,
\end{equation}
where $\bar{M}$ and $\bar{Q}$ represent the total mass and charge of
the outer geometry, respectively. At the boundary, the following
constraints are obtained from the continuity of the metric
coefficients of the metrics \eqref{g6} and \eqref{g20} as
\begin{eqnarray}\label{g21}
g_{tt}&=&e^{B\mathcal{H}^2+C}=1-\frac{2\bar{M}}{\mathcal{H}}+\frac{\bar{Q}^2}{\mathcal{H}^2},\\\label{g21a}
g_{rr}&=&e^{-A\mathcal{H}^2}=1-\frac{2\bar{M}}{\mathcal{H}}+\frac{\bar{Q}^2}{\mathcal{H}^2},\\\label{g22}
\frac{\partial g_{tt}}{\partial
r}&=&B\mathcal{H}e^{B\mathcal{H}^2+C}=\frac{\bar{M}}{\mathcal{H}^2}-\frac{\bar{Q}^2}{\mathcal{H}^3}.
\end{eqnarray}
The Krori-Barua unknowns ($A,B,C$) are calculated by solving
Eqs.\eqref{g21}-\eqref{g22} simultaneously as
\begin{eqnarray}\label{g23}
A&=&-\frac{1}{\mathcal{H}^2}
\ln\left(1-\frac{2\bar{M}}{\mathcal{H}}+\frac{\bar{Q}^2}{\mathcal{H}^2}\right),\\\label{g24}
B&=&\frac{1}{\mathcal{H}^2}\left(\frac{\bar{M}}{\mathcal{H}}-\frac{\bar{Q}^2}{\mathcal{H}^2}\right)
\left(1-\frac{2\bar{M}}{\mathcal{H}}+\frac{\bar{Q}^2}{\mathcal{H}^2}\right)^{-1},\\\label{g25}
C&=&\ln\left(1-\frac{2\bar{M}}{\mathcal{H}}+\frac{\bar{Q}^2}{\mathcal{H}^2}\right)
-\left(\frac{\bar{M}}{\mathcal{H}}-\frac{\bar{Q}^2}{\mathcal{H}^2}\right)
\left(1-\frac{2\bar{M}}{\mathcal{H}}+\frac{\bar{Q}^2}{\mathcal{H}^2}\right)^{-1}.
\end{eqnarray}
The acceptability of the above equations produces the following
restraint on the considered structure \cite{41i}
\begin{align}\label{g25a}
\frac{\left(\frac{\bar{M}}{\mathcal{H}}-\frac{\bar{Q}^2}{\mathcal{H}^2}\right)}{\left(1-\frac{2\bar{M}}{\mathcal{H}}
+\frac{\bar{Q}^2}{\mathcal{H}^2}\right)}+\left(\frac{2+\frac{2\bar{M}}{\mathcal{H}}-\frac{\bar{Q}^2}{\mathcal{H}^2}}{4
-\frac{2\bar{M}}{\mathcal{H}}+\frac{\bar{Q}^2}{\mathcal{H}^2}}\right)\ln\left(1-\frac{2\bar{M}}{\mathcal{H}}
+\frac{\bar{Q}^2}{\mathcal{H}^2}\right)=0.
\end{align}
The pressure in radial direction inside stellar systems must
disappear at the boundary ($r=\mathcal{H}$). Therefore, we obtain
from Eq.\eqref{g17} (in which $P_r$ shows radial pressure in
modified gravity) by combining it with Eqs.\eqref{g23}-\eqref{g25}
as
\begin{align}\nonumber
P_r|_{(r=\mathcal{H})}&=\bigg[\mathcal{H}^4\big(\mathcal{H}(\mathcal{H}-2\bar{M})+\bar{Q}^2\big)\bigg\{\frac{1}{\mathcal{H}(\mathcal{H}-2\bar{M})
+\bar{Q}^2}\big\{5\varpi\big(\mathcal{H}(3\mathcal{H}-4\bar{M})\\\nonumber
&+\bar{Q}^2\big)\ln\bigg(\frac{\bar{Q}^2-2\bar{M}\mathcal{H}+\mathcal{H}^2}{\mathcal{H}^2}\bigg)\big\}
-\varpi\ln\bigg(\frac{\bar{Q}^2-2\bar{M}\mathcal{H}+\mathcal{H}^2}{\mathcal{H}^2}\bigg)^2\\\nonumber
&+\frac{1}{\big(\mathcal{H}(\mathcal{H}-2\bar{M})+\bar{Q}^2\big)^2}\big\{32\pi\mathcal{H}^4\big(\mathcal{H}(\mathcal{H}-2\bar{M})
+\bar{Q}^2\big)+\varpi\big(\bar{Q}^2\mathcal{H}(47\\\nonumber
&\times\bar{M}-19\mathcal{H})+3\bar{M}\mathcal{H}^2(5\mathcal{H}-9\bar{M})-16\bar{Q}^4\big)\big\}\bigg\}
\bigg]^{-1}\bigg[\bar{Q}^2\big(\varpi\bar{M}\mathcal{H}+\mathcal{H}^4\\\nonumber
&\times(20\varpi\mathfrak{B_c}-2)\big)-\ln\bigg(\frac{\bar{Q}^2-2\bar{M}\mathcal{H}
+\mathcal{H}^2}{\mathcal{H}^2}\bigg)\big\{\bar{Q}^2\mathcal{H}\big(-2\varpi\bar{M}+2\mathcal{H}^3\\\nonumber
&+\varpi\mathcal{H}\big)+2\mathcal{H}^5(-2\bar{M}(2\varpi\mathfrak{B_c}+1)+4\varpi\mathfrak{B_c}\mathcal{H}+\mathcal{H})
+\varpi\bar{Q}^4\big\}-\varpi\bar{Q}^4\\\label{g26}
&+2\mathcal{H}^5\big(-8\varpi\mathfrak{B_c}\bar{M}+\bar{M}-16\pi\mathfrak{B_c}\mathcal{H}^3\big)\bigg]=0,
\end{align}
which provides the value of bag constant as
\begin{align}\label{g27}
\mathfrak{B_c}=\frac{\big(\varpi\bar{Q}^2+2\mathcal{H}^4\big)\bigg\{\bar{M}\mathcal{H}-\bar{Q}^2-\big(\mathcal{H}(\mathcal{H}-2\bar{M})
+\bar{Q}^2\big)\ln\bigg(\frac{\bar{Q}^2-2\bar{M}\mathcal{H}+\mathcal{H}^2}{\mathcal{H}^2}\bigg)\bigg\}}{4\mathcal{H}^4
\left\{2\varpi\mathcal{H}(\mathcal{H}-\bar{M})\ln\bigg(\frac{\bar{Q}^2-2\bar{M}\mathcal{H}+\mathcal{H}^2}{\mathcal{H}^2}\bigg)
+4\varpi\bar{M}\mathcal{H}-5\varpi\bar{Q}^2+8\pi\mathcal{H}^4\right\}}.
\end{align}

The experimental data of five different strange stars has been used
to calculate the unknowns $A,~B,~C$ and $\mathfrak{B_c}$. We also
determine the value of $\bar{Q}$ by utilizing the constraint
\eqref{g25a}. It is observed that these strange bodies are
compatible with the limit suggested by Buchdhal \cite{42a}, i.e.,
$\frac{2\bar{M}}{\mathcal{H}}<\frac{8}{9}$. The bag constant is
evaluated for two values of the coupling parameter as $\varpi=\pm5$
through which we analyze the stellar evolution successfully. The
observed data such as radii and masses of considered candidates are
presented in Table $\mathbf{1}$ and corresponding values of
Krori-Barua constants as well as $\frac{\bar{Q}^2}{\mathcal{H}^2}$
are given in Table $\mathbf{2}$. Tables $\mathbf{3}$ and
$\mathbf{4}$ provide the values of $\mathfrak{B_c}$ for each star
candidate with the constant $\varpi=5$ and $-5$, respectively.

The bag constant for strange stars having negligible mass has a
range of $58.9-91.5~MeV/fm^3$ \cite{42a1}. On the other hand, this
range becomes $56-78~MeV/fm^3$ for the quarks with a mass of
approximately $150~MeV$ \cite{42a2}. However, the larger values of
$\mathfrak{B_c}$ have also been proposed in the literature. Xu
\cite{42a3} suggested that a particular compact star, namely LMXB
EXO 0748-676 can be treated as the strange candidate for
$\mathfrak{B_c}=60~MeV/fm^3$ and $110~MeV/fm^3$. Further, CERN-SPS
and RHIC have released some experimental findings according to which
the bag model depending on density may provide a wide range of the
values of this constant \cite{41i,42a4}. Some specific values of the
bag constant (such as $83,~100$ and $120~MeV/fm^3$) have been used
by Deb \emph{et al.} \cite{37a} to construct a strange quark model.
Remarkably, the values of $\mathfrak{B_c}$ for different strange
stars are obtained for $\varpi=5$ as $97.89,~54.79,~191.43,~226.34$
and $111.14$ $MeV/fm^3$, while $\varpi=-5$ provides these values as
$97.58,~54.78,~190.75,~226.13$ and $111.12$ $MeV/fm^3$. Since we
have done the whole analysis in a non-minimally coupled theory, the
presence of an extra force offers larger values of the bag constant
for some star candidates. However, the bag constant for 4U 1820-30,
Vela X-I and Her X-I are found to be within the allowable range
discussed above.

\section{Graphical Investigation of Compact Stars like Structures}

Here, we discuss several features of all the compact structures
which are associated with anisotropic fluid distribution in
$f(\mathcal{R},\mathcal{T},\mathcal{Q})$ framework. To check the
physical acceptance of developed solution \eqref{g16}-\eqref{g18}
for $\varpi=\pm5$, the state variables are graphically analyzed for
each stellar candidate by using their respective experimental as
well as calculated data as given in Tables $\mathbf{1}$ and
$\mathbf{2}$. We examine the behavior of metric functions,
anisotropic pressure, energy conditions, mass, redshift and
compactness. We also discuss their stability for both values of the
model parameter. It is known that the singularity-free and
positively increasing behavior of the metric components in the whole
domain ensures the compatibility of the solution. Equation
\eqref{g15} demonstrates the metric coefficients in terms of
Krori-Barua constants whose calculated values are given in Table
$\mathbf{2}$. The graphs of both potentials are plotted in Figure
$\mathbf{1}$ which show that the resulting solution is physically
consistent. It should be mentioned that the green color shows RXJ
1856-37 star, blue expresses Vela X-I, red signifies Her X-I, yellow
indicates 4U 1820-30 and black represents SAX J 1808.4-3658 in all
plots.
\begin{table}[H]
\scriptsize \centering \caption{Preliminary data of five different
stars \cite{38}} \label{Table1} \vspace{+0.1in}
\setlength{\tabcolsep}{0.95em}
\begin{tabular}{cccccc}
\hline\hline Star Models & 4U 1820-30 & Vela X-I & SAX J 1808.4-3658
& RXJ 1856-37 & Her X-I
\\\hline $Mass(M_{\bigodot})$ & 2.25 & 1.77 & 1.435 & 0.9041 & 0.88
\\\hline
$\mathcal{H}(km)$ & 10 & 12.08 & 7.07 & 6 & 7.7
\\\hline
$\frac{\bar{M}}{\mathcal{H}}$ & 0.331 & 0.215 & 0.298 & 0.222 & 0.168  \\
\hline\hline
\end{tabular}
\end{table}
\begin{table}[H]
\scriptsize \centering \caption{Values of triplet ($A,B,C$) for
different stars} \label{Table2} \vspace{+0.1in}
\setlength{\tabcolsep}{0.95em}
\begin{tabular}{cccccc}
\hline\hline Star Models & 4U 1820-30 & Vela X-I & SAX J 1808.4-3658
& RXJ 1856-37 & Her X-I
\\\hline $A$ & 0.0095062 & 0.0038254 & 0.0169186 & 0.0160570 & 0.0068809
\\\hline
$B$ & 0.0073157 & 0.0025435 & 0.0126937 & 0.0107699 & 0.0042356
\\\hline
$C$ & -1.68219 & -0.92939 & -1.48017 & -0.96577 & -0.65909
\\\hline
$\frac{\bar{Q}^2}{\mathcal{H}^2}$ & 0.048 & 0.003 & 0.026 & 0.004 & 0.001  \\
\hline\hline
\end{tabular}
\end{table}
\begin{table}[H]
\scriptsize \centering \caption{Value of state parameters and bag
constant for different stars for $\varpi=5$} \label{Table3}
\vspace{+0.1in} \setlength{\tabcolsep}{0.95em}
\begin{tabular}{cccccc}
\hline\hline Star Models & 4U 1820-30 & Vela X-I & SAX J 1808.4-3658
& RXJ 1856-37 & Her X-I
\\\hline $\mathfrak{B_c}$ ($MeV/fm^3$) & 0.000129550 & 0.000072511 & 0.000253342 & 0.000299542 & 0.000147080
\\\hline
$\mu_c (gm/cm^3)$ & 1.5104$\times$10$^{15}$ & 6.0589$\times$10$^{14}$ & 2.6943$\times$10$^{15}$ & 2.5445$\times$10$^{15}$ &
1.0849$\times$10$^{15}$
\\\hline
$\mu_s (gm/cm^3)$ & 6.9405$\times$10$^{14}$ & 4.4134$\times$10$^{14}$ & 7.7459$\times$10$^{14}$ & 8.2061$\times$10$^{14}$ &
6.3666$\times$10$^{14}$
\\\hline
$P_{rc} (dyne/cm^2)$ & 2.4589$\times$10$^{35}$ & 6.5063$\times$10$^{34}$ & 4.0125$\times$10$^{35}$ & 2.7944$\times$10$^{35}$ &
8.8691$\times$10$^{34}$  \\
\hline\hline
\end{tabular}
\end{table}
\begin{table}[H]
\scriptsize \centering \caption{Value of state parameters and bag
constant for different stars for $\varpi=-5$} \label{Table4}
\vspace{+0.1in} \setlength{\tabcolsep}{0.95em}
\begin{tabular}{cccccc}
\hline\hline Star Models & 4U 1820-30 & Vela X-I & SAX J 1808.4-3658
& RXJ 1856-37 & Her X-I
\\\hline $\mathfrak{B_c}$ ($MeV/fm^3$) & 0.000129143 & 0.000072497 & 0.000252436 & 0.000299263 & 0.000147056
\\\hline
$\mu_c (gm/cm^3)$ & 1.4221$\times$10$^{15}$ & 5.5652$\times$10$^{14}$ & 2.5258$\times$10$^{15}$ & 2.3304$\times$10$^{15}$ &
9.8569$\times$10$^{14}$
\\\hline
$\mu_s (gm/cm^3)$ & 5.9479$\times$10$^{14}$ & 3.9452$\times$10$^{14}$ & 5.9479$\times$10$^{14}$ & 6.1779$\times$10$^{14}$ &
5.4877$\times$10$^{14}$
\\\hline
$P_{rc} (dyne/cm^2)$ & 3.3091$\times$10$^{35}$ & 8.7151$\times$10$^{34}$ & 5.4446$\times$10$^{35}$ & 3.7984$\times$10$^{35}$&
1.1957$\times$10$^{35}$  \\
\hline\hline
\end{tabular}
\end{table}

\subsection{Study of Physical Variables}

In self-gravitating stellar bodies, the state variables such as
energy density, radial and tangential pressures should be maximum at
their core. For each star candidate, the plots of these variables
with respect to the model \eqref{g61} are shown in Figure
$\mathbf{2}$ which assure the existence of extremely dense objects
as all variables gain their maximum values at the center ($r=0$)
inside the stellar systems. The radial pressure insider each
candidate vanishes at the boundary for both values of $\varpi$,
while the other two variables show monotonically decreasing behavior
with rise in $r$, as presented in Figure $\mathbf{2}$. Tables
$\mathbf{3}$ and $\mathbf{4}$ provide the values of state
determinants at the core as well as boundary of each star for
$\varpi=5$ and $-5$, respectively. It is seen that all considered
compact structures become more dense for $\varpi=5$ but the pressure
components take higher values for $\varpi=-5$. Figure $\mathbf{3}$
shows the regular behavior of energy density and radial pressure as
$\frac{d\mu}{dr} < 0$ and $\frac{dP_r}{dr} < 0$. Therefore, the
graphical interpretation guarantees the existence of highly compact
anisotropic matter configurations in this theory.
\begin{figure}\center
\epsfig{file=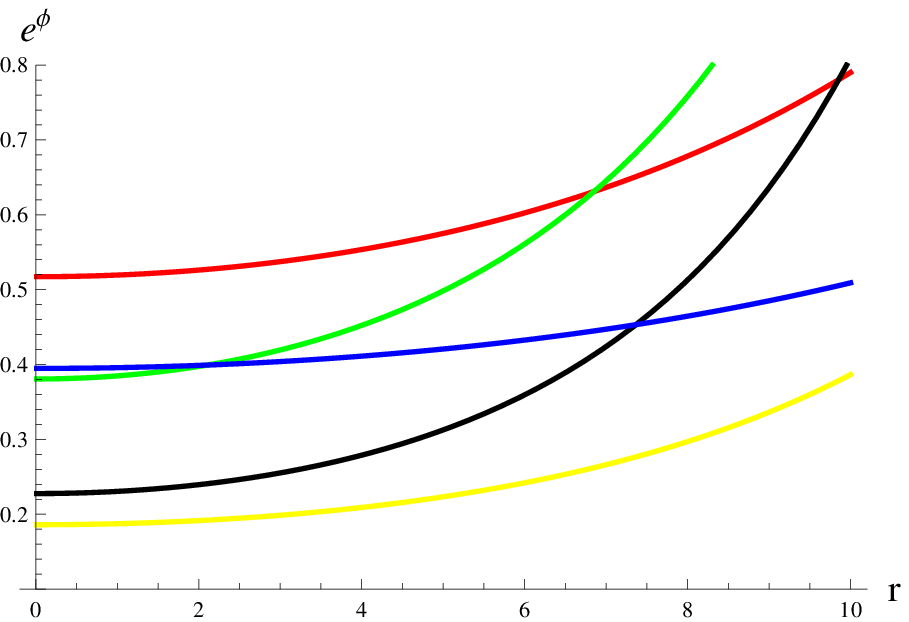,width=0.45\linewidth}\epsfig{file=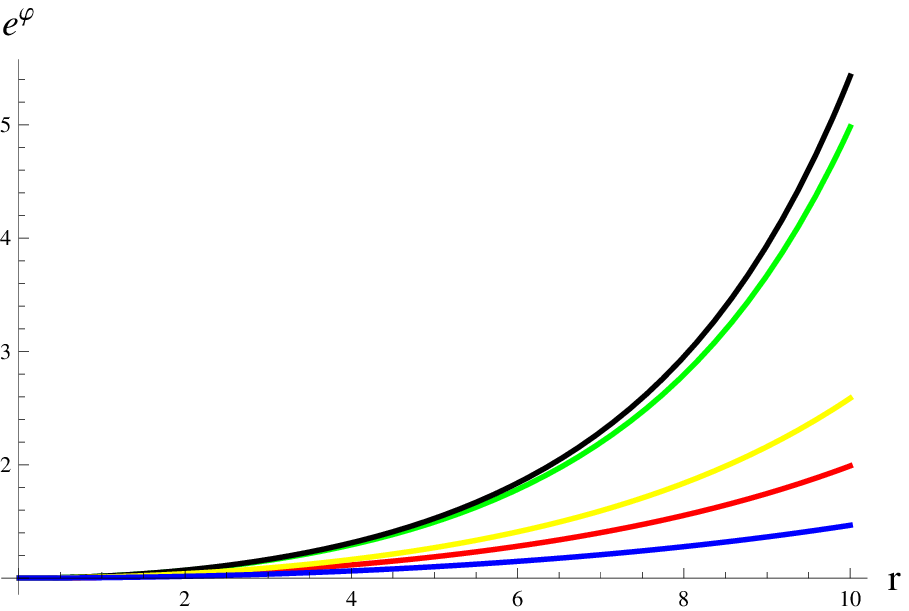,width=0.45\linewidth}
\caption{Plots of temporal/radial metric functions versus $r$}
\end{figure}
\begin{figure}\center
\epsfig{file=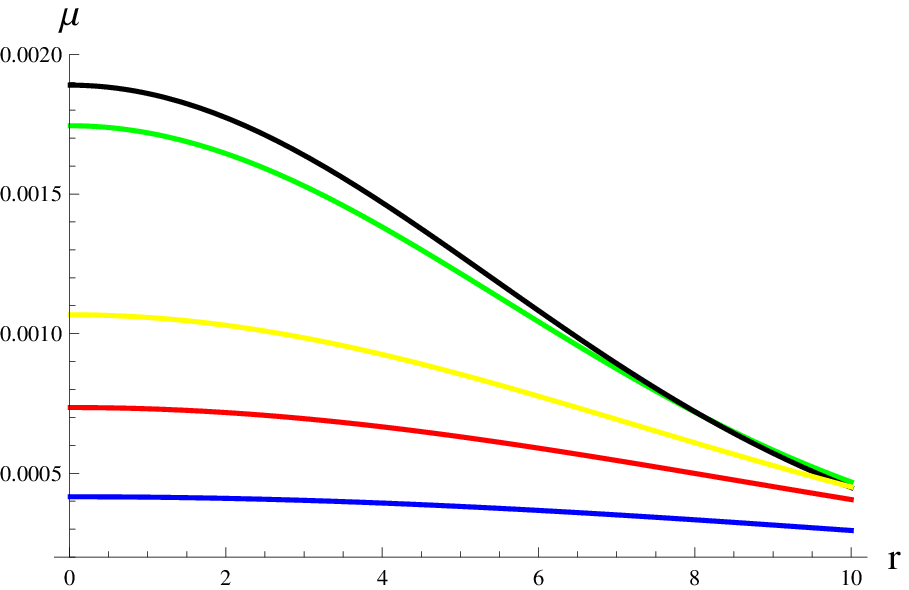,width=0.45\linewidth}\epsfig{file=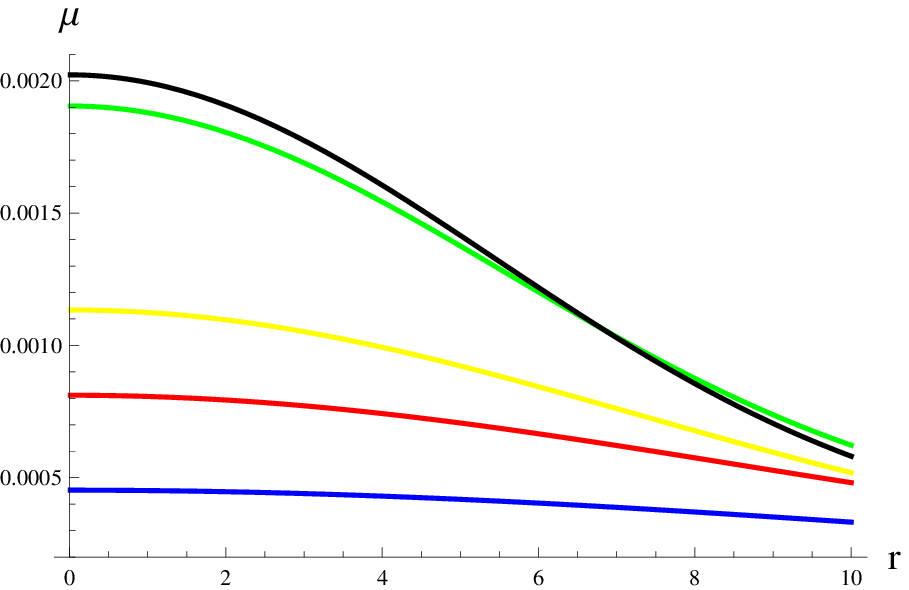,width=0.45\linewidth}
\epsfig{file=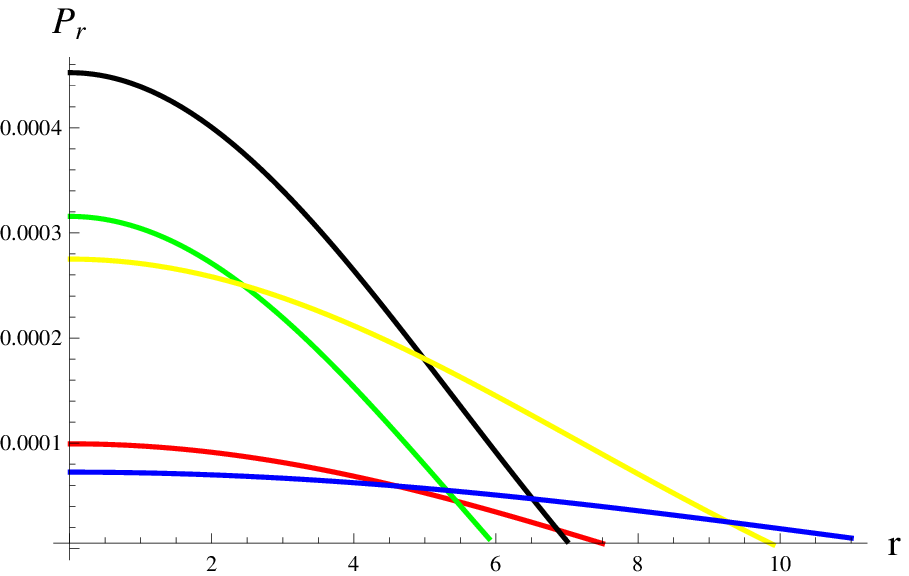,width=0.45\linewidth}\epsfig{file=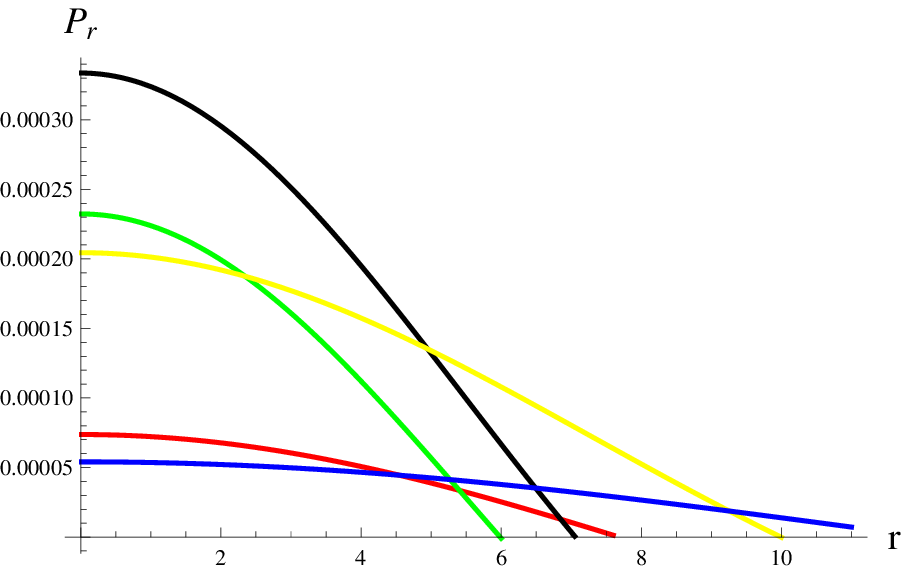,width=0.45\linewidth}
\epsfig{file=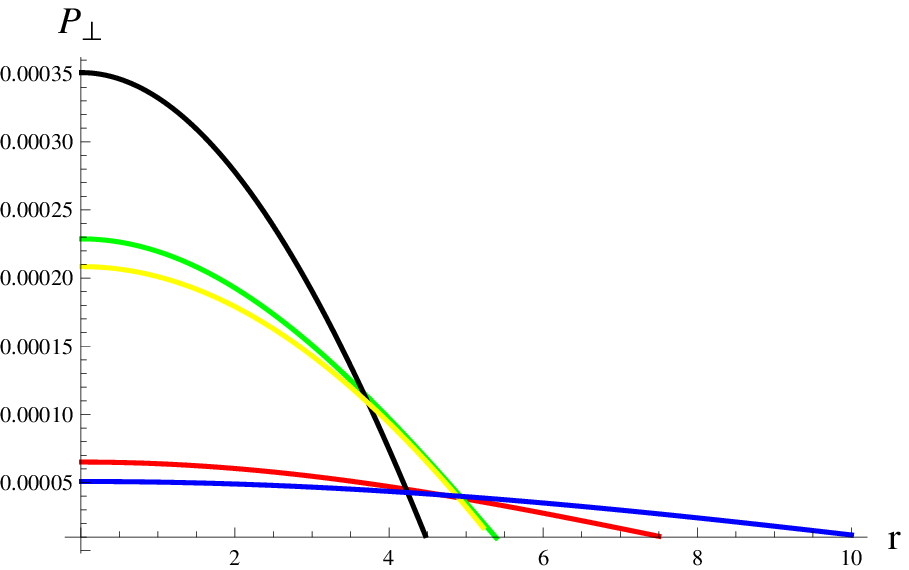,width=0.45\linewidth}\epsfig{file=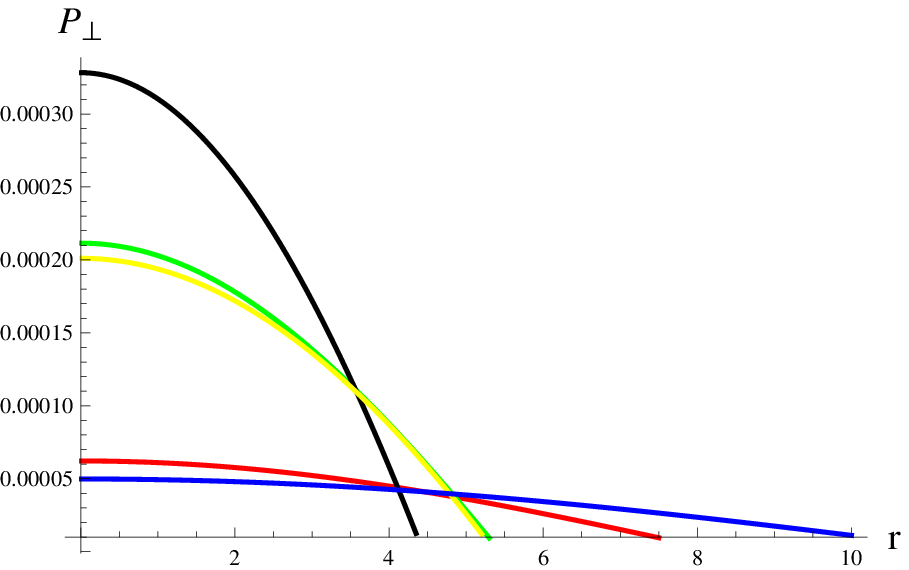,width=0.45\linewidth}
\caption{Plots of state determinants versus $r$ corresponding to
$\varpi=-5$ (left) and $\varpi=5$ (right)}
\end{figure}

\subsection{Effect of Anisotropic Pressure}

The expression for anisotropic pressure (i.e., $\Delta=P_\bot-P_r$)
can be calculated by using Eqs.\eqref{g17} and \eqref{g18} as
\begin{align}\nonumber
\Delta&=\frac{1}{4}e^{-Ar^2}\bigg[4\big(-A(1+Br^2)+B(2+Br^2)\big)\bigg\{8\pi+\varpi
e^{-Ar^2}\bigg(3B+\frac{1-e^{Ar^2}}{r^2}\\\nonumber
&+B^2r^2-A(2+Br^2)\bigg)\bigg\}^{-1}+\frac{2q^2}{r^4}\big(\varpi\big(A+ABr^2-B\big(2+Br^2\big)\big)-2e^{Ar^2}\big)\\\nonumber
&+16r^4\varpi e^{Ar^2}\bigg\{\frac{1}{4r^2}\big((A+B)
\big(3A^2r^2+A\big(Br^2-2\big)+2B\big(Br^2-2\big)\big) \big(\varpi
q^2\\\nonumber
&+2r^4\big)\big)+8\pi\mathfrak{B_c}e^{Ar^2}\big(A\big(2+Br^2\big)-2B-3A^2r^2\big)
+\frac{\varpi\mathfrak{B_c}}{2r^2}\big(A^3r^4\big(11Br^2\\\nonumber
&+12\big)-A^2r^2\big(-6e^{Ar^2}+7B^2r^4+52Br^2+14\big)+B\big(4e^{Ar^2}+3B^3r^6-4\\\nonumber
&-10B^2r^4-21Br^2\big)+A\big(-2e^{Ar^2}\big(Br^2+2\big)+B^3r^6+18B^2r^4+55Br^2\\\nonumber
&+4\big)\big)\bigg\}\bigg\{\big(\varpi\big(-Ar^2\big(Br^2+2\big)+B^2r^4+3Br^2+1\big)+e^{Ar^2}\big(8\pi
r^2-\varpi\big)\big)\\\nonumber
&\times\big(\varpi\big(-A^2r^4-5Ar^2\big(2Br^2+3\big)+3B^2r^4+23Br^2+4\big)+4e^{Ar^2}\big(8\pi
r^2\\\nonumber &-\varpi\big)\big)
\bigg\}^{-1}\bigg]-\bigg\{r^2\big\{\varpi\big(-A^2r^4-5Ar^2\big(2Br^2+3\big)+3B^2r^4+23Br^2+4\big)\\\nonumber
&-4e^{Ar^2}\big(\varpi-8\pi
r^2\big)\big\}\bigg\}^{-1}\bigg\{A\big(8\varpi\mathfrak{B_c}Br^6+\varpi
q^2+r^4(8\varpi\mathfrak{B_c}
+2)\big)-4\mathfrak{B_c}r^2\\\label{g19}
&\times\big(\varpi+e^{Ar^2}\big(8\pi r^2-\varpi
\big)\big)+B\big(\varpi
q^2+r^4(2-24\varpi\mathfrak{B_c})\big)\bigg\}.
\end{align}
We examine the effect of anisotropy on the structural development of
considered stars by means of experimental findings (presented in
Table $\mathbf{1}$) and Krori-Barua constants. The anisotropic
pressure exerts in outward and inward direction according to the
case when $P_\bot>P_r$ (which yields $\Delta>0$) and $P_\bot<P_r$
(i.e., $\Delta<0$), respectively. For a viable
$f(\mathcal{R},\mathcal{T},\mathcal{Q})$ model \eqref{g61}, Figure
$\mathbf{4}$ shows the behavior of anisotropy corresponding to each
candidate. It is observed that $\Delta$ shows monotonically
increasing behavior towards the boundary and remains positive
throughout for all stars which confirms the existence of repelling
force due to which structural evolution in massive geometries
happens.
\begin{figure}\center
\epsfig{file=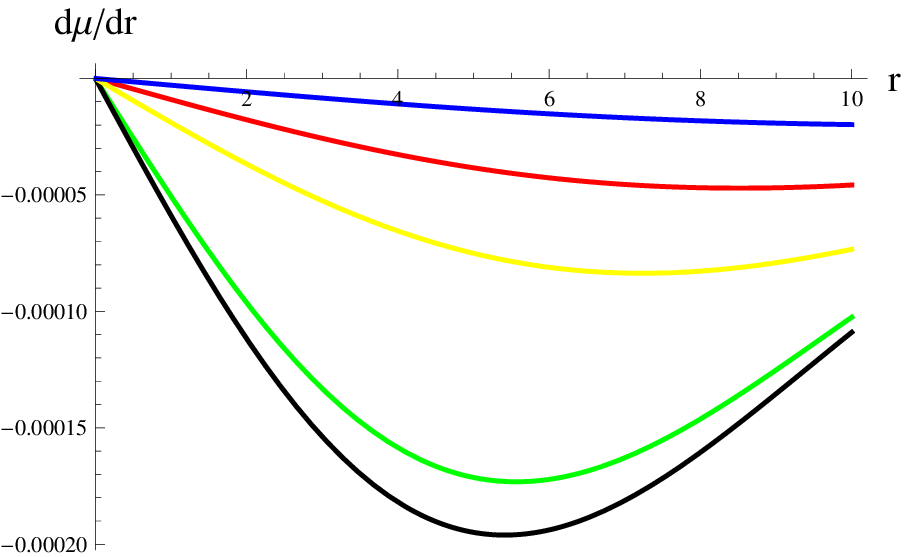,width=0.45\linewidth}\epsfig{file=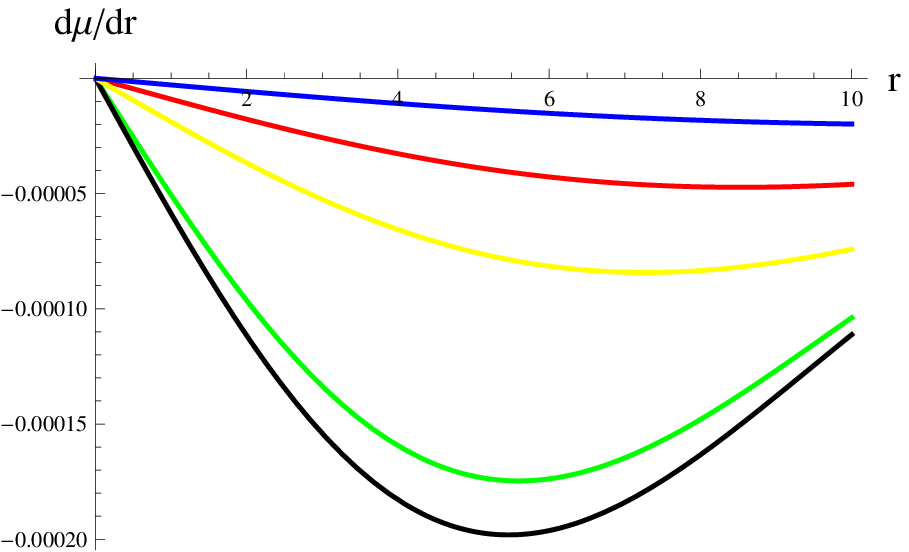,width=0.45\linewidth}
\epsfig{file=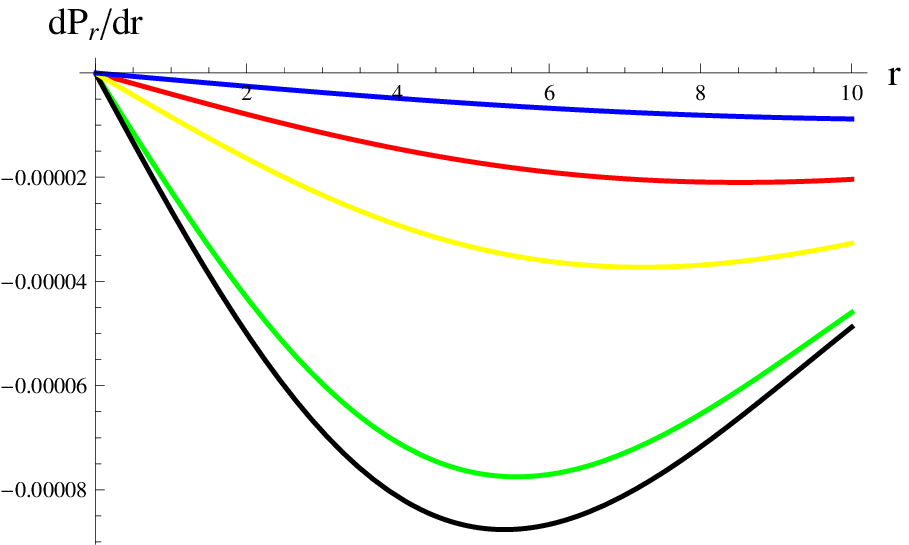,width=0.45\linewidth}\epsfig{file=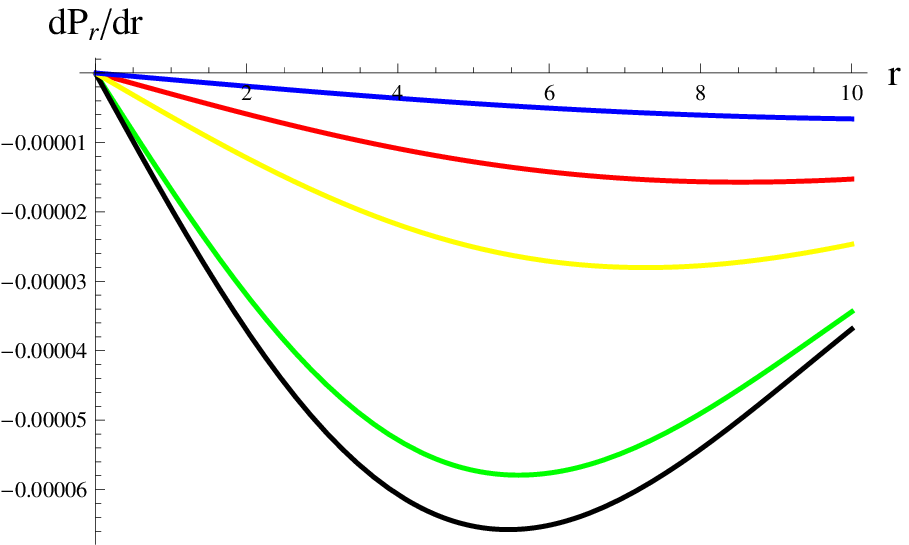,width=0.45\linewidth}
\caption{Plots of $\frac{d\mu}{dr}$ and $\frac{dP_r}{dr}$ versus $r$
corresponding to $\varpi=-5$ (left) and $\varpi=5$ (right)}
\end{figure}
\begin{figure}\center
\epsfig{file=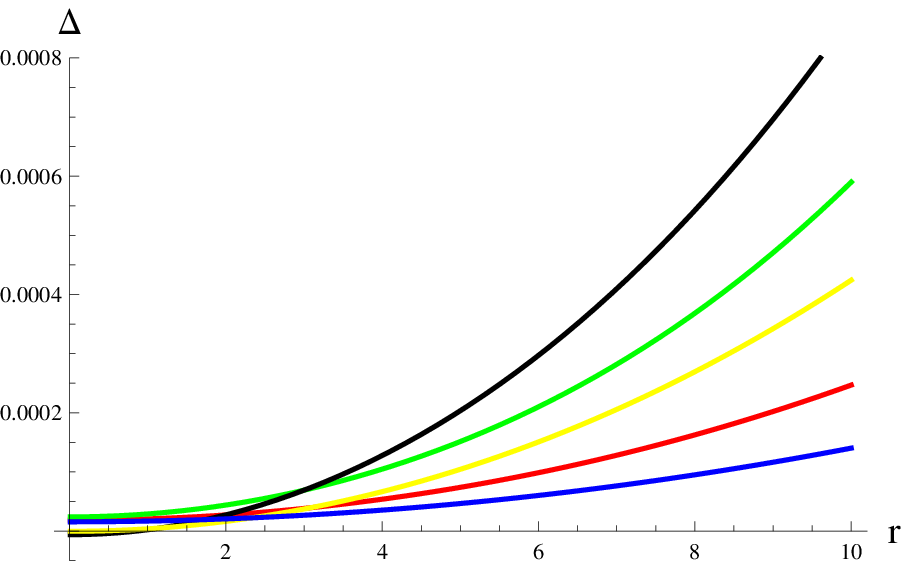,width=0.45\linewidth}\epsfig{file=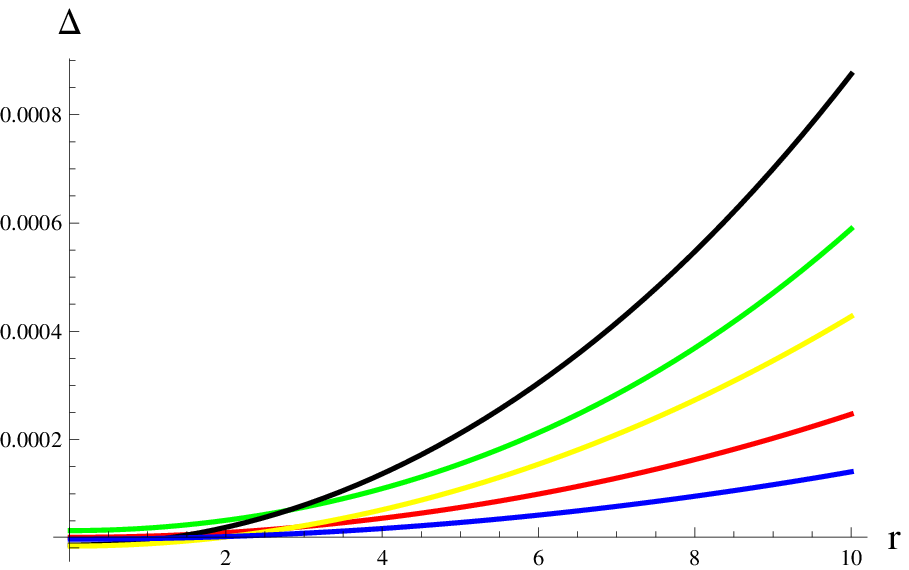,width=0.45\linewidth}
\epsfig{file=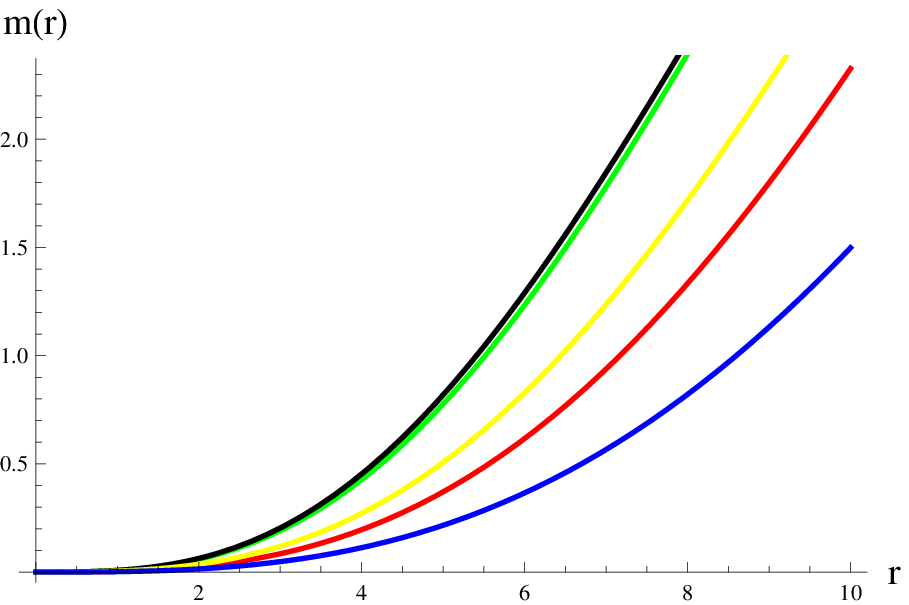,width=0.45\linewidth}\epsfig{file=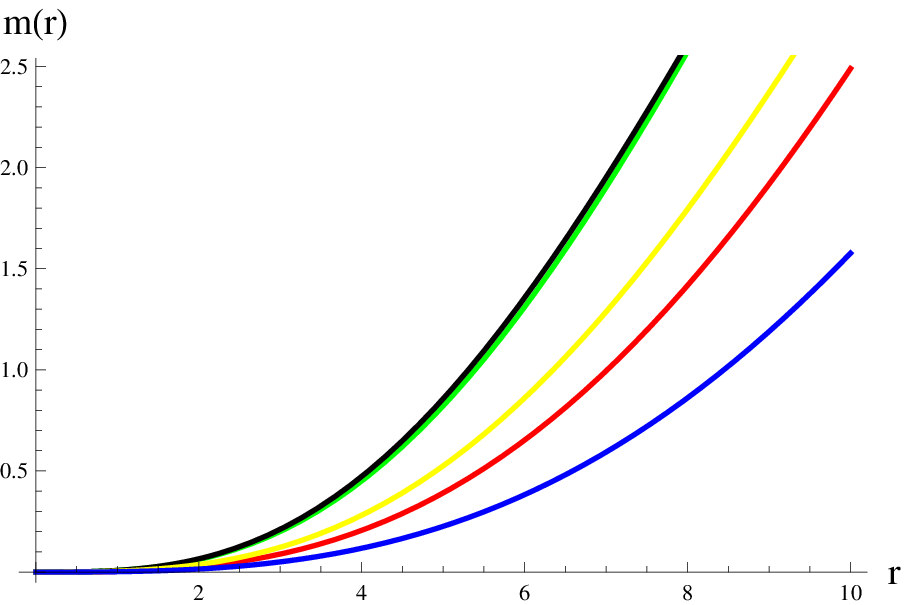,width=0.45\linewidth}
\caption{Plots of anisotropic factor and mass versus $r$
corresponding to $\varpi=-5$ (left) and $\varpi=5$ (right)}
\end{figure}

\subsection{Mass, Redshift and Compactness Parameters}

The mass of inner self-gravitating systems can be characterized as
\cite{42aa}
\begin{equation}\label{g63}
m(r)=4\pi\int_{0}^{\mathcal{H}}r^2\mu dr,
\end{equation}
where Eq.\eqref{g16} exhibits the value of effective energy density
($\mu$). We employ an initial condition $m(0)=0$ to obtain numerical
solution of Eq.\eqref{g63} for our considered model and examine the
graphical behavior of mass inside each compact star, as shown in
Figure $\mathbf{4}$. Various physical properties of astronomical
objects can be examined to study their structural composition. The
compactness factor $\big(\sigma(r)\big)$ is one of them which
describes the ratio of mass and radius of a compact star. Buchdahl
\cite{42a} proposed the upper limit of $\sigma(r)$ by matching the
inner and outer spacetimes at hypersurface ($r=\mathcal{H}$) and
concluded that the system will remain stable as long as its value is
less than $\frac{4}{9}$.

The study of interior redshift of a compact body helps to describe
the density profile of that body. The traveling distance of a moving
photon from the core of a star to its surface becomes longer as
photon must travel through much denser core. This ultimately results
in the dispersion and loss in energy of those photons. On the other
hand, a photon travels shorter distance and moves through a less
dense region when it comes out from near the surface. Consequently,
there happens less dispersion and less energy losses. However, the
mathematical expression for surface redshift in terms of compactness
can be given as
\begin{equation}
D(r)=\frac{1}{\sqrt{1-2\sigma(r)}}-1.
\end{equation}
For a feasible compact star coupled with perfect matter
distribution, Buchdahl established the value of $D(r)$ to be less
than $2$, whereas this limit was observed as $5.211$ by Ivanov
\cite{42b} in the case of anisotropic configuration when the
dominant energy condition is satisfied. Both the above factors for
each quark candidate are plotted in Figure $\mathbf{5}$, from which
we observe them within their acceptable ranges for both values of
$\varpi$. Moreover, these factors take less values at the surface
for $\varpi=-5$.
\begin{figure}\center
\epsfig{file=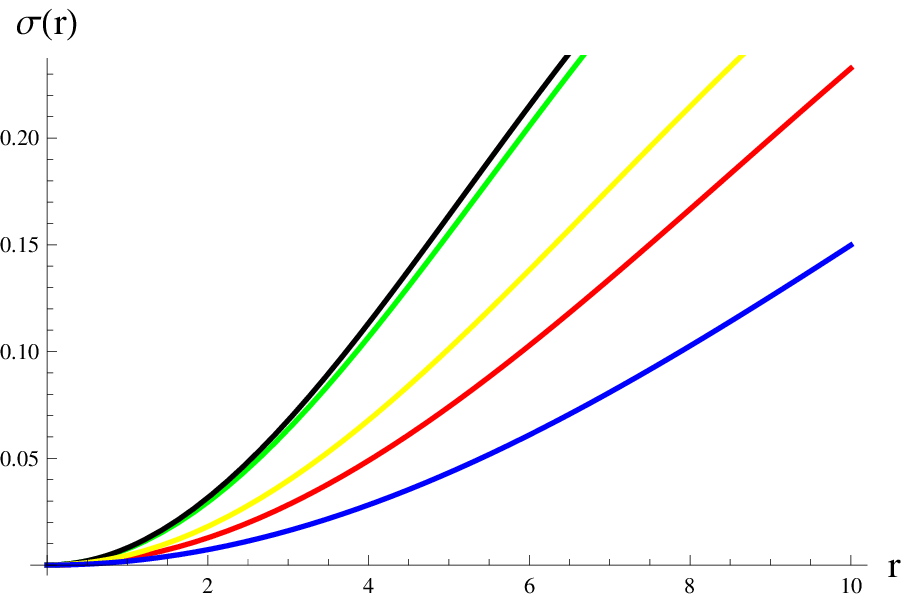,width=0.45\linewidth}\epsfig{file=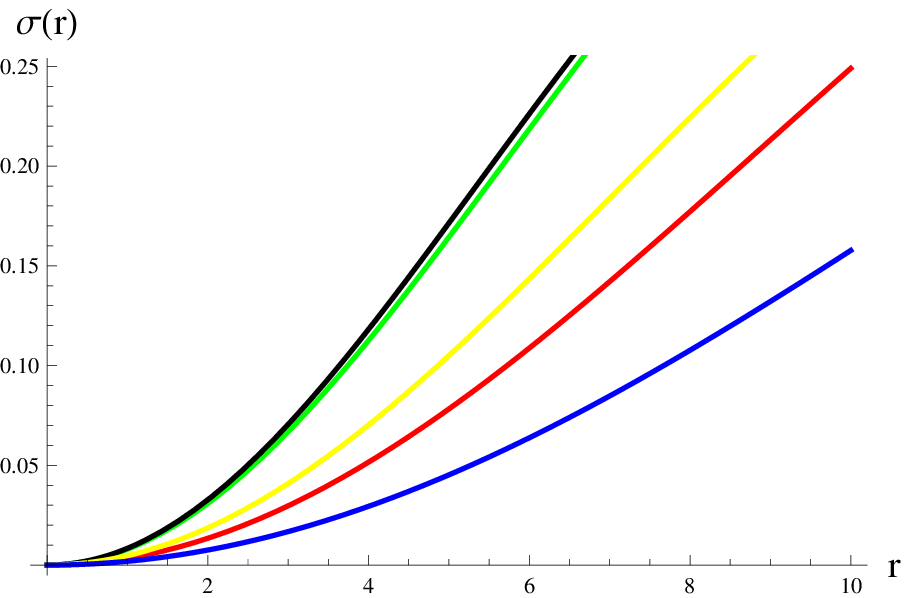,width=0.45\linewidth}
\epsfig{file=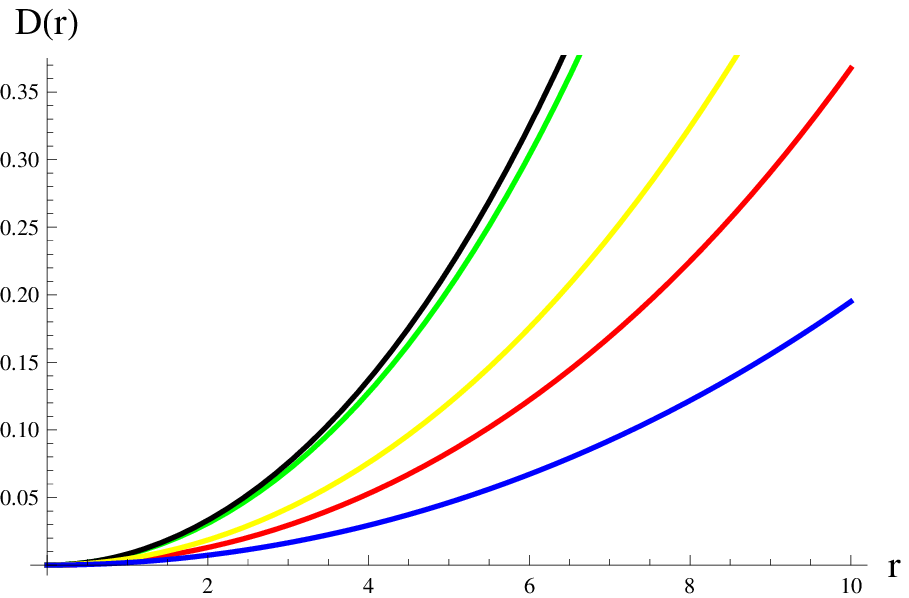,width=0.45\linewidth}\epsfig{file=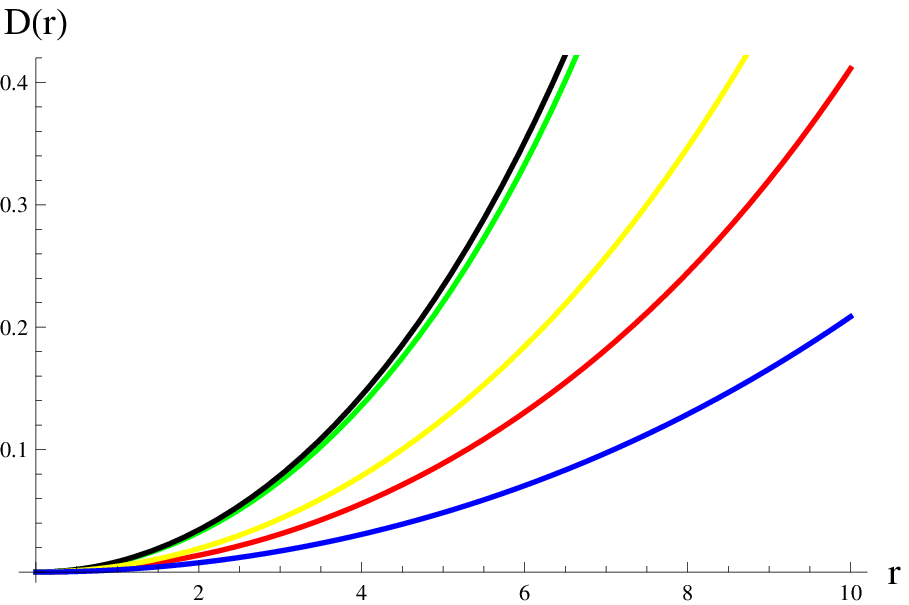,width=0.45\linewidth}
\caption{Plots of compactness and redshift factors versus $r$
corresponding to $\varpi=-5$ (left) and $\varpi=5$ (right)}
\end{figure}

\subsection{Energy Conditions}

In astrophysics, some constraints (energy conditions) attain much
significance which are used to determine the nature of matter
distribution in a stellar body. The fulfilment of such conditions
helps to distinguish the exotic or usual matter in the interior of a
self-gravitating geometry. The viability of a resulting solution to
the field equations can also be studied by means of these bounds.
Energy conditions must be satisfied by the state variables
satisfying ordinary matter with a particular geometry. For charged
anisotropic matter distribution in
$f(\mathcal{R},\mathcal{T},\mathcal{Q})$ theory, they are given in
the form
\begin{eqnarray}\nonumber
&&\mu+\frac{q^2}{8\pi r^4} \geq 0, \quad \mu+P_r \geq 0,\\\nonumber
&&\mu+P_\bot+\frac{q^2}{4\pi r^4} \geq 0, \quad
\mu-P_r+\frac{q^2}{4\pi r^4} \geq 0,\\\label{g50} &&\mu-P_\bot \geq
0, \quad \mu+P_r+2P_\bot+\frac{q^2}{4\pi r^4} \geq 0.
\end{eqnarray}
Figures $\mathbf{6}$ and $\mathbf{7}$ present the plots of all the
above conditions for $\varpi=5$ and $-5$, respectively, whose
positive trend in the whole domain confirms the viability of
$f(\mathcal{R},\mathcal{T},\mathcal{Q})$ model \eqref{g61} and the
obtained solution. As a result, all quark candidates must contain
normal matter in their interiors.
\begin{figure}\center
\epsfig{file=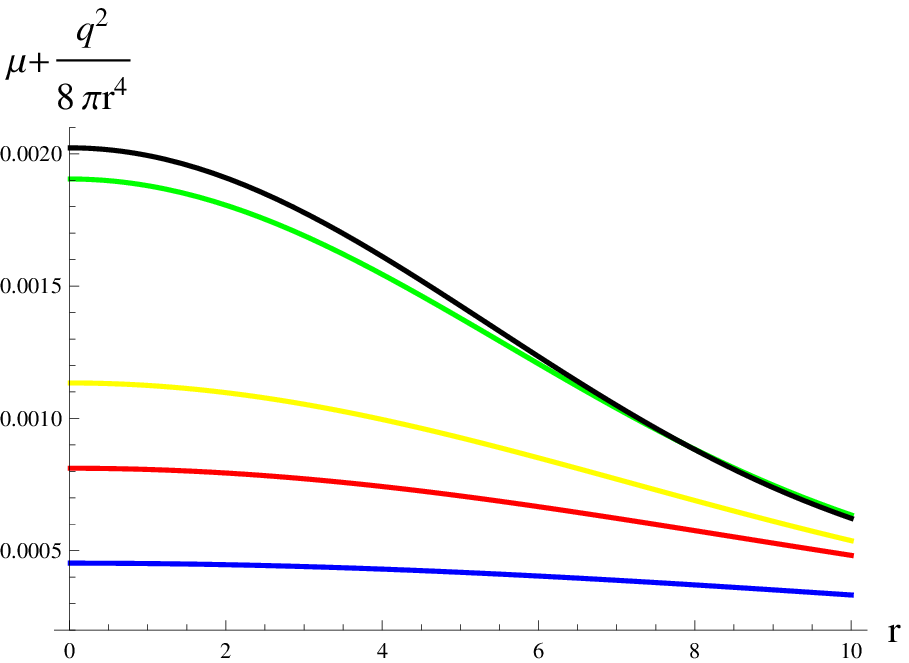,width=0.45\linewidth}\epsfig{file=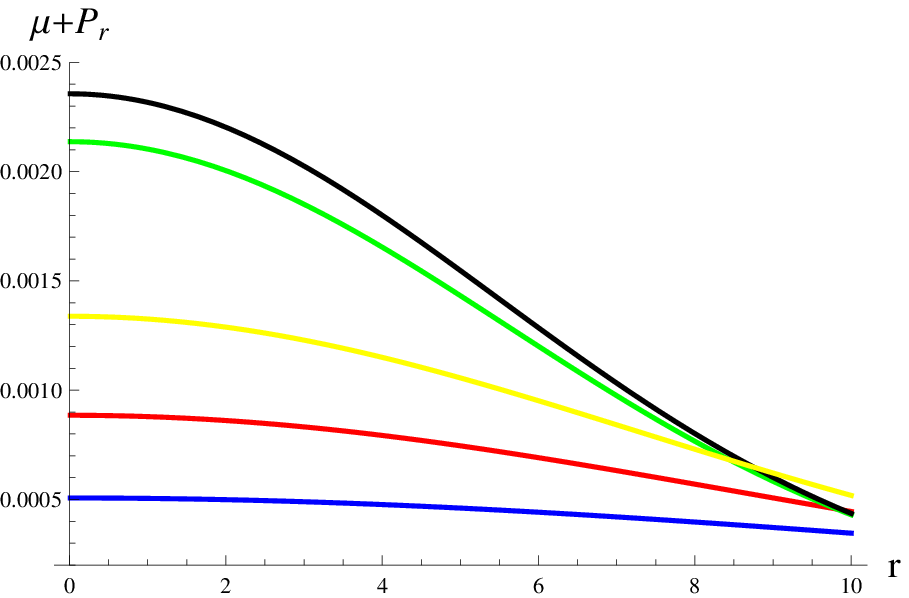,width=0.45\linewidth}
\epsfig{file=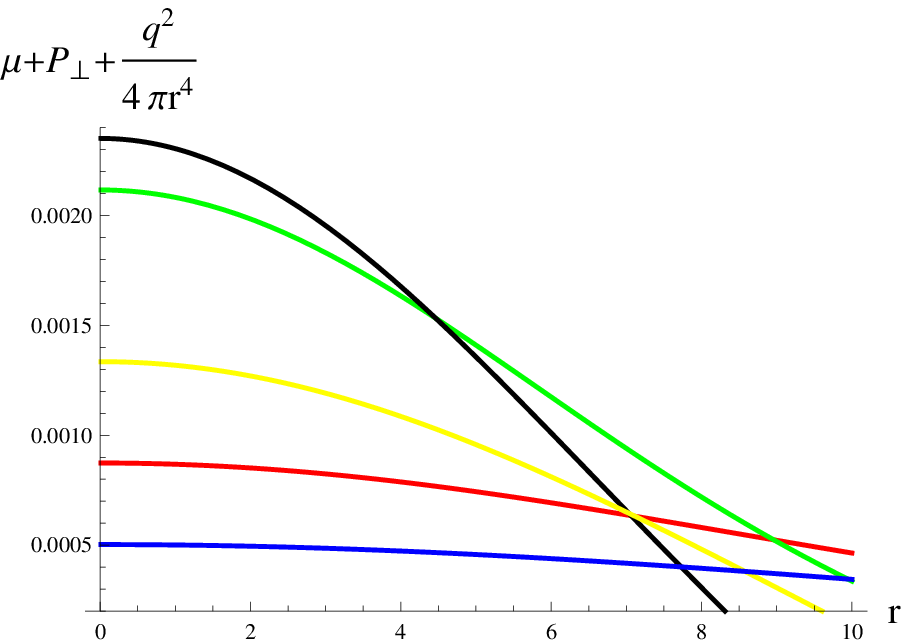,width=0.45\linewidth}\epsfig{file=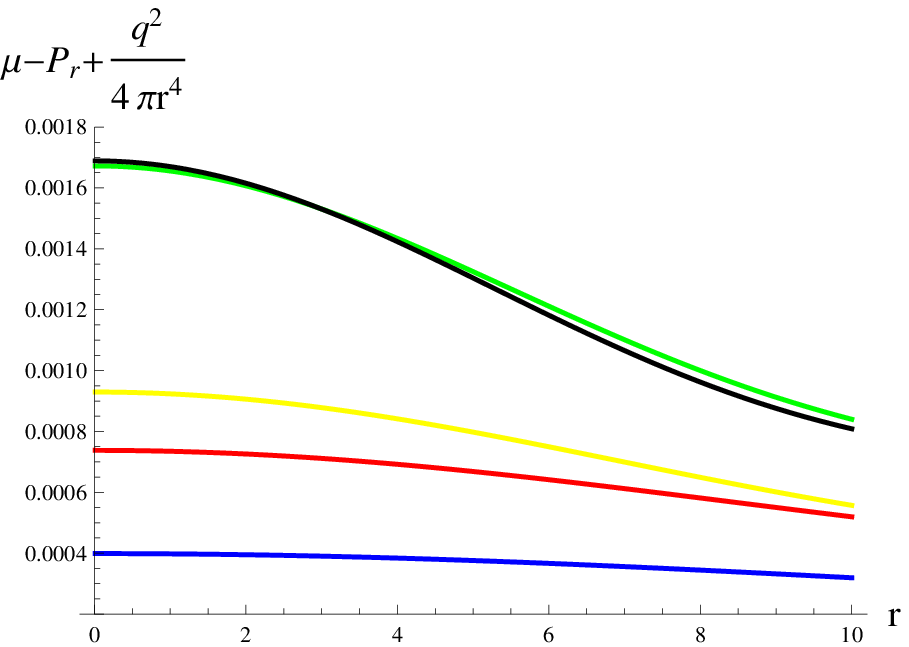,width=0.45\linewidth}
\epsfig{file=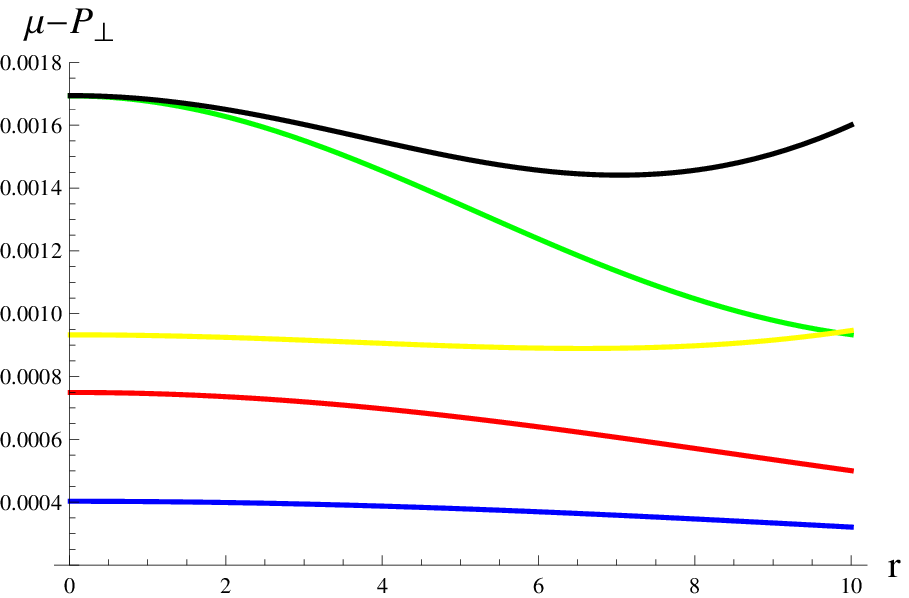,width=0.45\linewidth}\epsfig{file=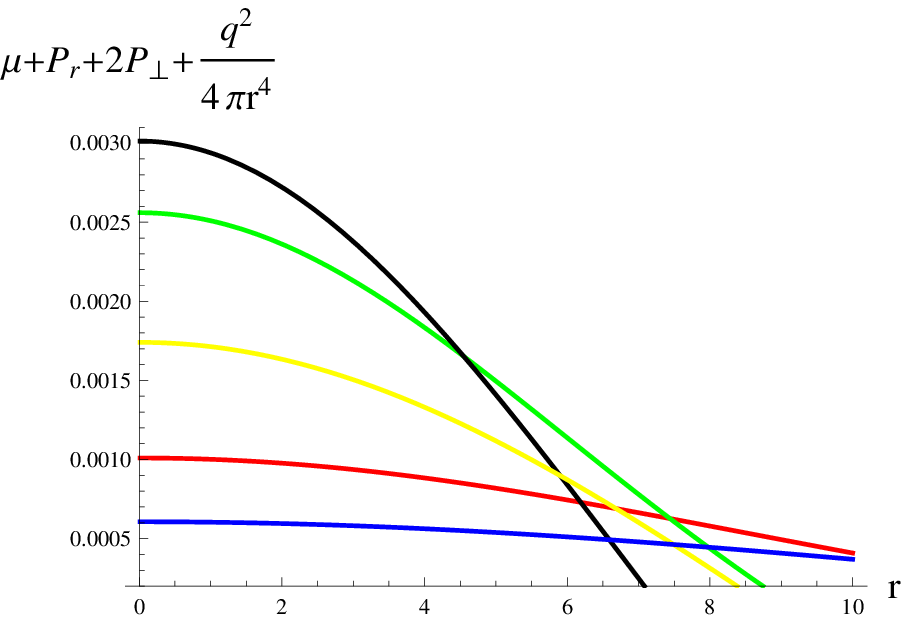,width=0.45\linewidth}
\caption{Plots of energy conditions versus $r$ corresponding to
$\varpi=5$}
\end{figure}
\begin{figure}\center
\epsfig{file=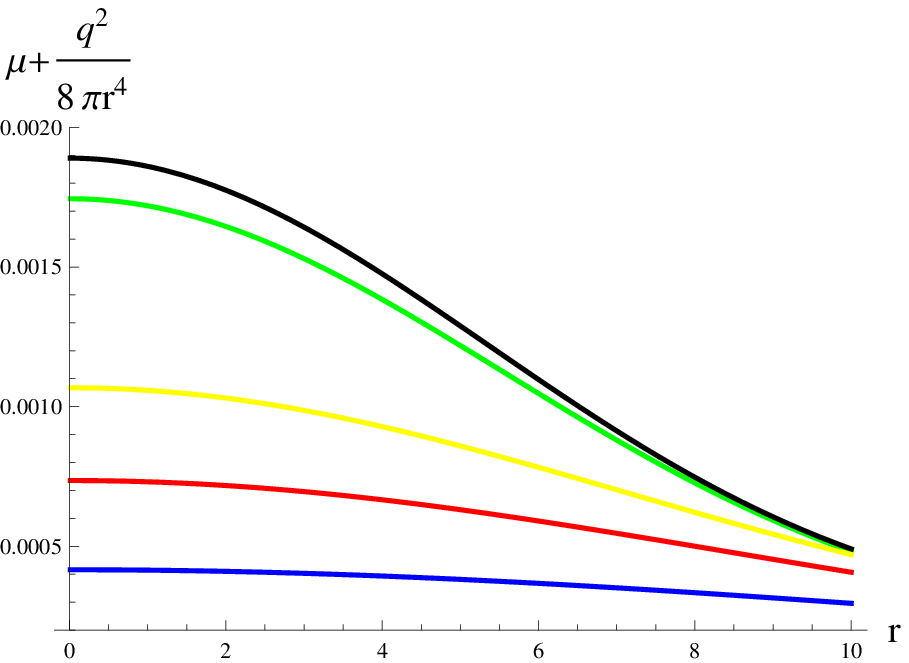,width=0.45\linewidth}\epsfig{file=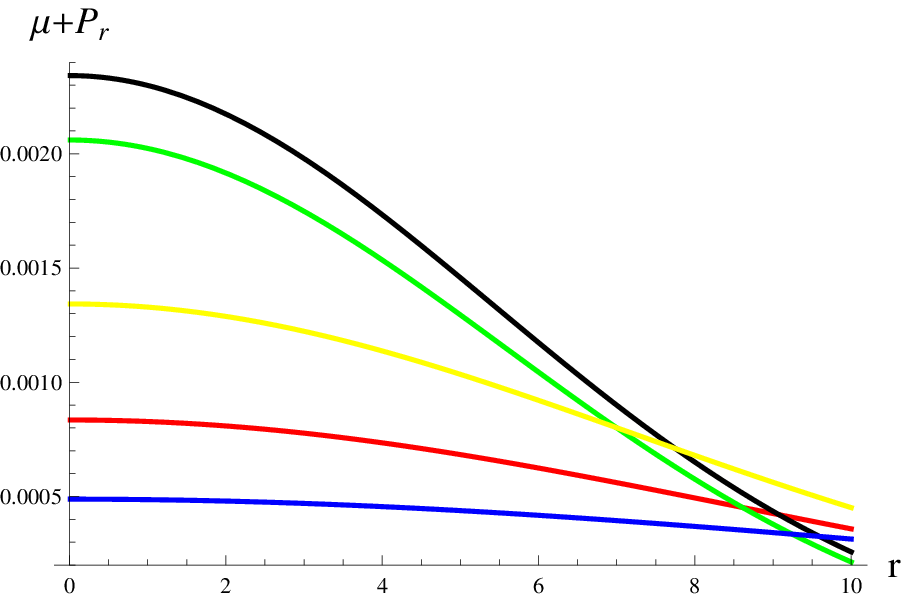,width=0.45\linewidth}
\epsfig{file=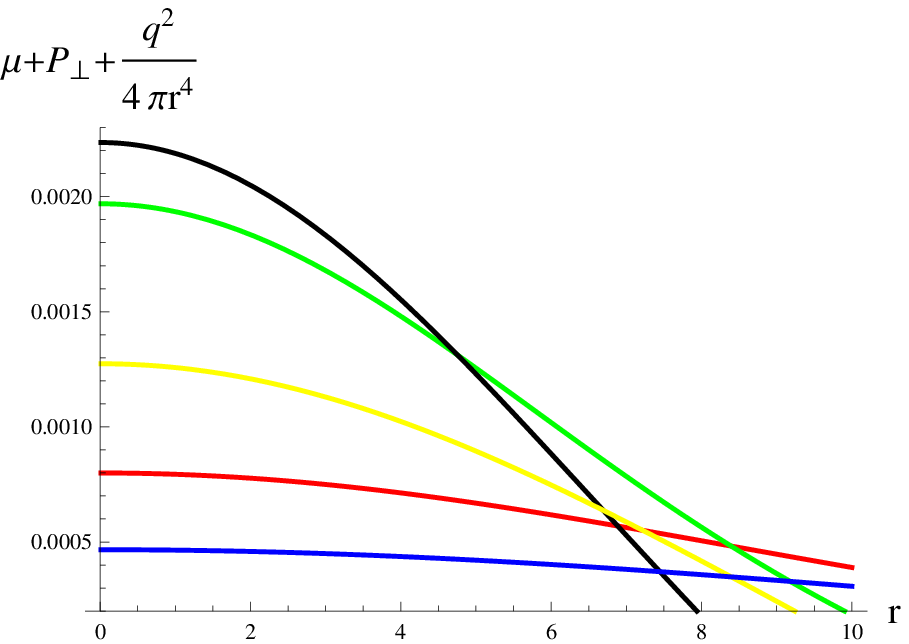,width=0.45\linewidth}\epsfig{file=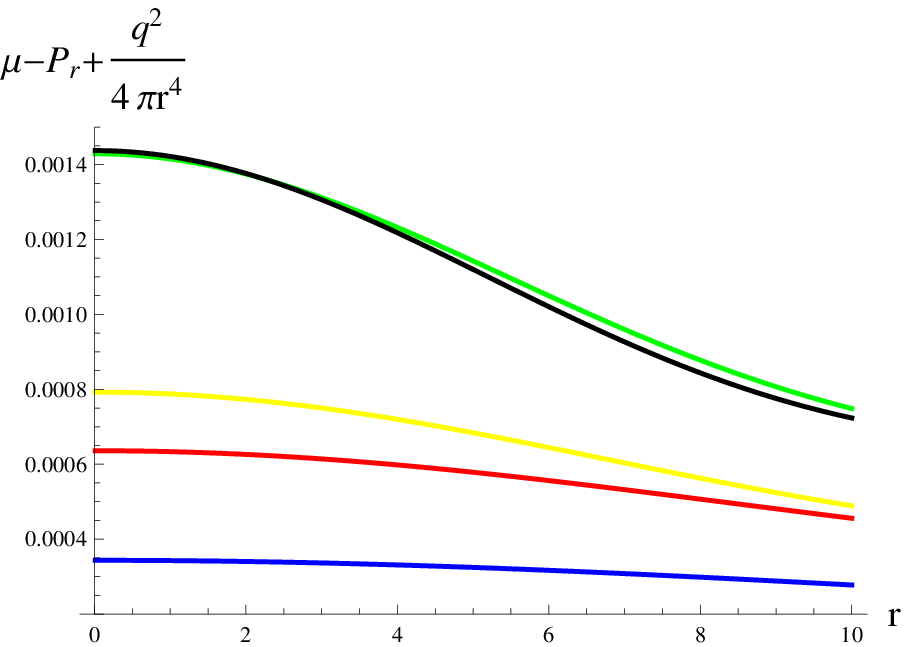,width=0.45\linewidth}
\epsfig{file=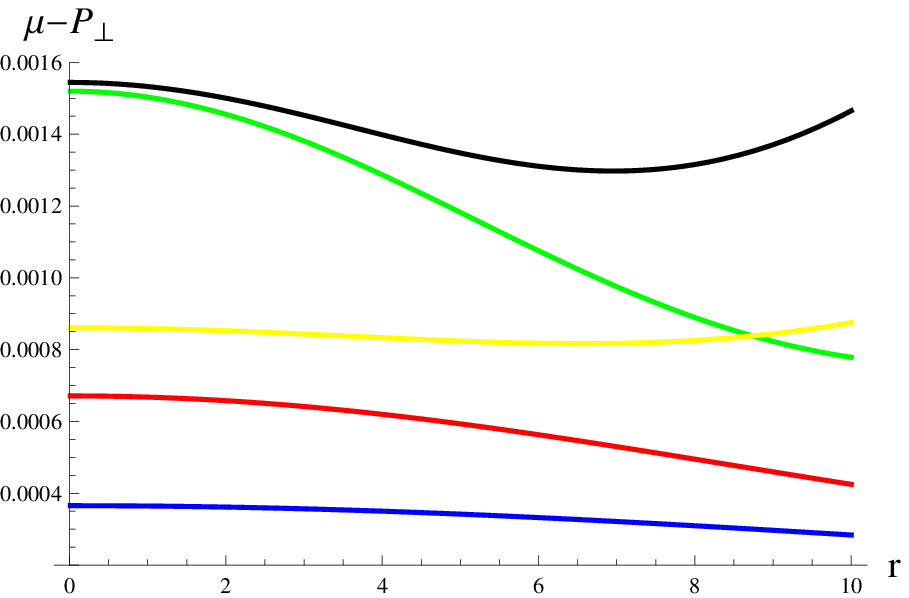,width=0.45\linewidth}\epsfig{file=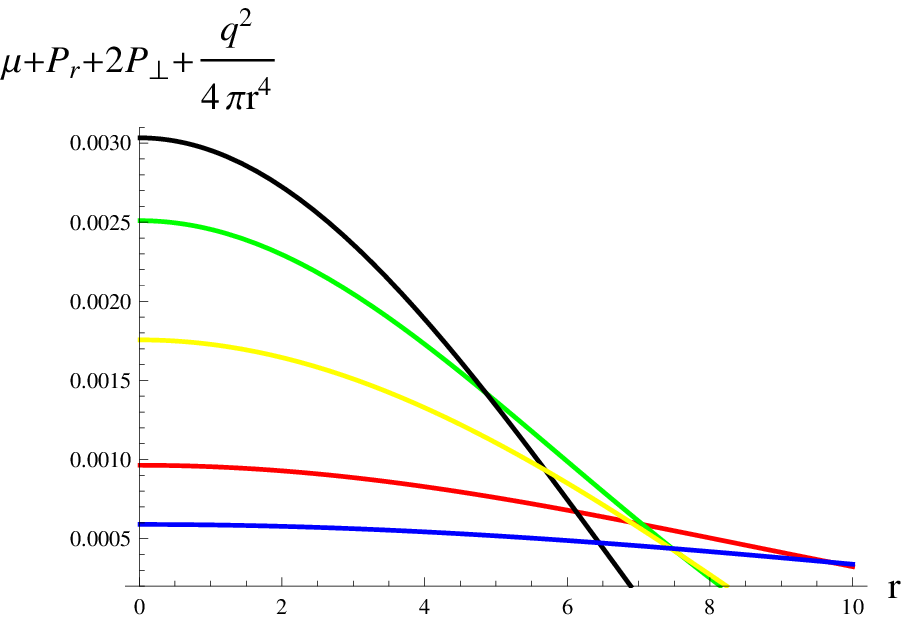,width=0.45\linewidth}
\caption{Plots of energy conditions versus $r$ corresponding to
$\varpi=-5$}
\end{figure}

\subsection{Analysis of Stable Regions}

In astronomy, the feasibility of physical models which explain
different eras of our universe depends on their stability. The
celestial systems which fulfill the stability criteria are more
appealing and hence the study of their structural development has
become a topic of great interest. Multiple approaches can be used to
examine the stability, two of them are used here to determine
whether the considered candidates are stable or not in this gravity.
One is the Herrera cracking concept \cite{42c} which is based on the
sound speed $v_{s}^{2}$. The value of squared sound speed,
$v_{s}^{2}=\frac{dP}{d\mu}$, to be within $(0,1)$, i.e., $0 <
v_{s}^{2}< 1$ inside the stellar object confirms its stability. This
is called causality condition. In anisotropic matter configuration,
there exist radial ($v_{sr}^{2}=\frac{dP_{r}}{d\mu}$) and tangential
components ($v_{s\bot}^{2}=\frac{dP_{\bot}}{d\mu}$) of sound speed
satisfying $0 < v_{sr}^{2} < 1$ and $0 < v_{s\bot}^{2} < 1$. Thus
the compact object shows stable behavior if the inequality $0 < \mid
v_{s\bot}^{2}-v_{sr}^{2} \mid < 1$ holds.

Another effective tool to investigate the stability of compact stars
is the adiabatic index $\big(\Gamma\big)$ whose value should be
greater than $\frac{4}{3}$ in the whole domain to obtain stable
models. This approach has widely been used in literature \cite{42d}.
Here, $\Gamma$ is characterized as
\begin{equation}\label{g62}
\Gamma=\frac{\mu+P_{r}}{P_{r}}
\left(\frac{dP_{r}}{d\mu}\right)=\frac{\mu+P_{r}}{P_{r}}
\left(v_{sr}^{2}\right).
\end{equation}
Figure $\mathbf{8}$ demonstrates the graphical behavior of $\mid
v_{s\bot}^{2}-v_{sr}^{2} \mid$ and $\Gamma$ for all candidates
corresponding to the calculated values of charge as well as bag
constant. The graphs are plotted for both values of coupling
constant. It is observed from the upper two plots that the quark
candidates show stable behavior throughout for $\varpi=5$. When
$\varpi=-5$, they are stable near their center but become unstable
towards the boundary.
\begin{figure}\center
\epsfig{file=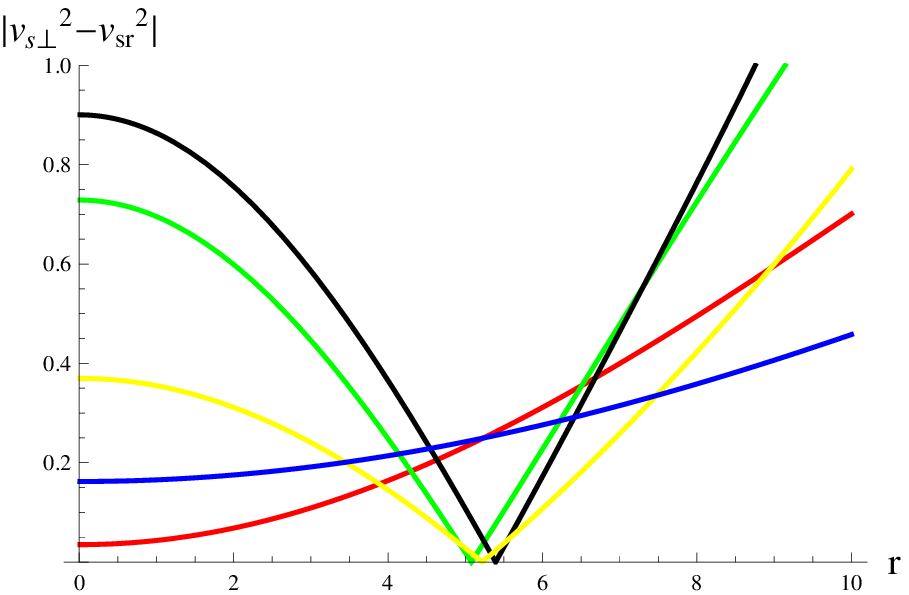,width=0.45\linewidth}\epsfig{file=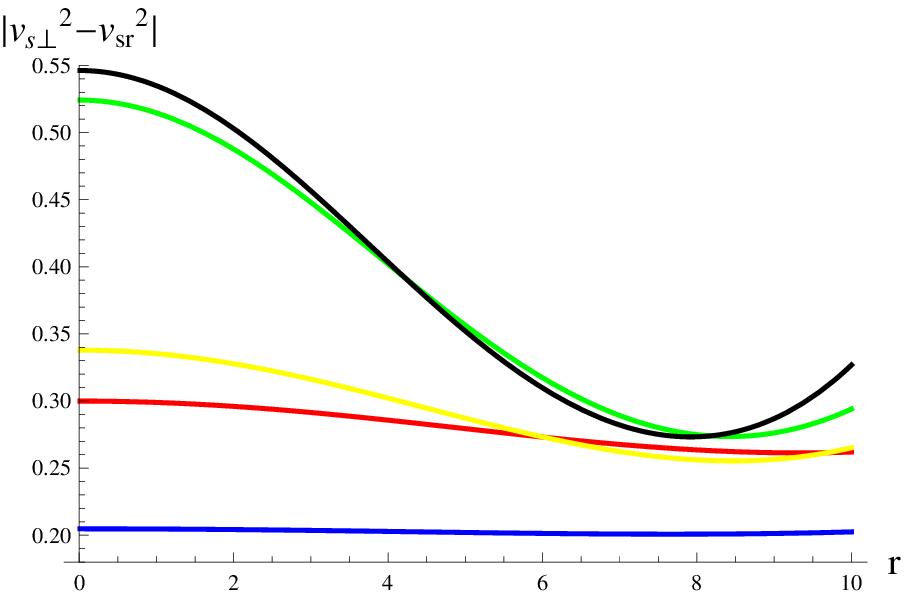,width=0.45\linewidth}
\epsfig{file=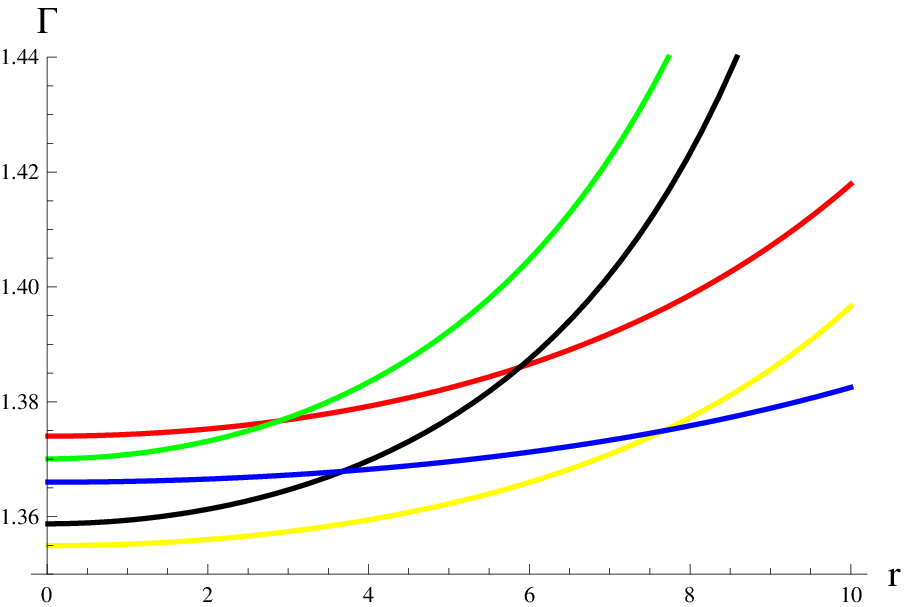,width=0.45\linewidth}\epsfig{file=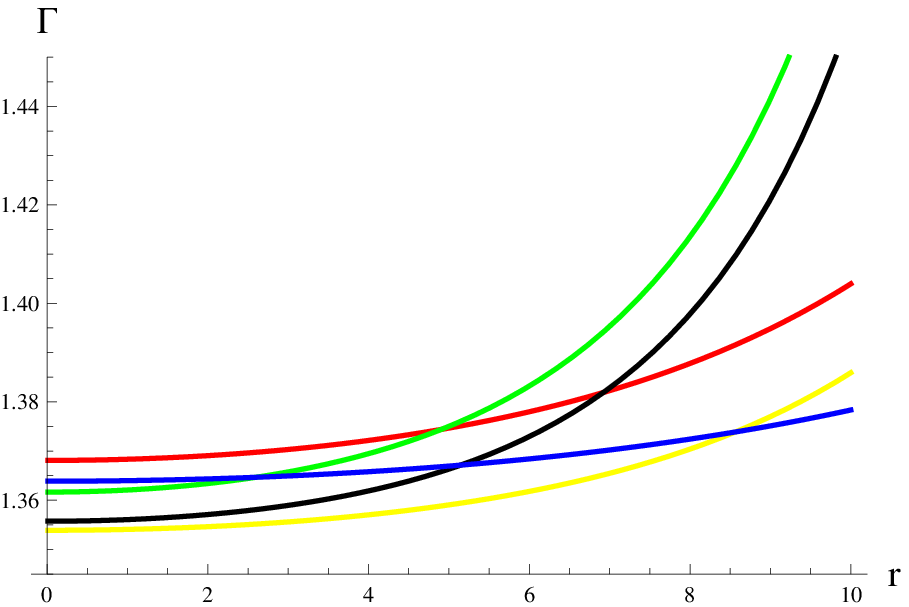,width=0.45\linewidth}
\caption{Plots of $\mid v_{s\bot}^{2}-v_{sr}^{2} \mid$ and adiabatic
index versus $r$ corresponding to $\varpi=-5$ (left) and $\varpi=5$
(right)}
\end{figure}

\section{Coupling Constant for Considered Quark Candidates}

The main objective of this article is to find the values of the bag
constant for different strange star candidates and then check the
physical acceptability of the resulting solution to the modified
field equations. We have achieved this from Eq.(36) where the bag
constant is explicitly expressed in terms of mass, radius and charge
of the compact star, and the coupling constant. However, we can also
explicitly express the coupling constant from the same equation to
find its values for different stars that give the observed values of
$\mathfrak{B_c}$. For this, we adopt $\mathfrak{B_c}=83~MeV/fm^3$
(which is within the allowed limit \cite{35}) to calculate the
values of $\varpi$ for all the considered candidates. By using the
preliminary data of quark stars (given in Table \textbf{1}), we find
the coupling constant as follows
\begin{itemize}
\item $\varpi=-1.12$ corresponding to a quark candidate 4U 1820-30.
\item $\varpi=+2.77$ corresponding to a quark candidate Vela X-I.
\item $\varpi=-5.02$ corresponding to a quark candidate SAX J
1808.4-3658.
\item $\varpi=-6.61$ corresponding to a quark candidate RXJ 1856-37.
\item $\varpi=-3.04$ corresponding to a quark candidate Her X-I.
\end{itemize}

\section{Conclusions}

This paper analyzes various physical attributes of some particular
anisotropic quark candidates (such as Vela X-I, 4U 1820-30, SAX J
1808.4-3658, Her X-I and RXJ 1856-37) which are influenced from
electromagnetic field by means of MIT bag model EoS in
$f(\mathcal{R},\mathcal{T},\mathcal{Q})$ gravity. We have adopted
$\varpi=\pm5$ along with a linear model as $\mathcal{R}+\varpi
\mathcal{Q}$ which preserves the strong non-minimal matter-geometry
interaction on massive particles. The complex gravitational
equations of motion along with TOV equation have been formulated by
making use of the bag model EoS \eqref{g14a}. We have evaluated the
values of $\frac{\bar{Q}^2}{\mathcal{H}^2}$ and bag constant
$\mathfrak{B_c}$ for each star candidate. Further, we have utilized
$g_{tt}$ and $g_{rr}$ metric components suggested by Krori-Barua
containing a triplet $(A,B,C)$ and calculated their values from
matching conditions (Table $\mathbf{2}$) after using the
observational data for each star. The solution to the field
equations has been obtained by making use of two particular
equations of state which interlink energy density and pressure
ingredients. The state variables as well as pressure anisotropy
indicate that the obtained solution shows physically acceptable
behavior (for instance, energy density and pressures are maximum at
the center and decreasing towards boundary) for both values of
$\varpi$.

The values of physical variables for $\varpi=\pm5$ have been
provided in Tables $\mathbf{3}$ and $\mathbf{4}$. It has been
observed that all quark stars become more dense for $\varpi=5$,
whereas radial and tangential pressures inside these structures gain
higher values for $\varpi=-5$. The graphical values of redshift and
compactness are found to be within their acceptable bounds. The
positive trend of all energy conditions for both cases have
confirmed that our developed solution is physically viable and
stellar bodies contain usual matter in their interiors. We have
analyzed the stability by using Herrera's cracking approach and
adiabatic index. For stable star, the inequality $0 < \mid
v_{s\bot}^{2}-v_{sr}^{2} \mid < 1$ must hold throughout the system.
It is seen that the considered quark candidates are stable
everywhere for $\varpi=5$, while two of them (RXJ 1856- 37 and SAX J
1808.4-3658) become unstable for $\varpi=-5$ (upper two plots of
Figure $\mathbf{8}$). However, it may be possible that some other
values of this constant provides unstable stars that are now stable
for both values of $\varpi$.

The graphical analysis for adiabatic index provides stable
structures in the whole domain for both values. The physical
variables such as energy density and pressure for two compact stars
(namely Her X-1 and RXJ 1856-37) have been calculated at their core
as well as the boundary in the context of GR \cite{37ag}. We compare
these values with the obtained results in modified theory and found
them lesser in this case. It should also be pointed out that a
strange candidate Her X-1 is observed to be denser in
$f(\mathcal{G})$ gravity than this theory \cite{38gg}. However, a
compact star SAX J 1808.4-3658 has a more dense interior in this
framework as compared to the $f(\mathcal{R},\mathcal{T})$ theory
\cite{25aa}. We conclude that
$f(\mathcal{R},\mathcal{T},\mathcal{Q})$ theory along with MIT EoS
\eqref{g14a} offers more suitable results for $\varpi=5$ in
comparison with \cite{25a,25b} and the solution for $\varpi=-5$.
Finally, all these results can be retrieved in GR by taking
$\varpi=0$ in modified model \eqref{g61}.

\section*{Appendix A}

The Krori-Barua triplet provides the adiabatic index as
\begin{align}\nonumber
\Gamma&=-\bigg[3 \big\{A \big(8 \varpi  \mathfrak{B_c}  B r^6+\varpi  q^2+r^4 (8 \varpi  \mathfrak{B_c} +2)\big)
-4 \mathfrak{B_c}r^2 \big(e^{A r^2} \big(8 \pi  r^2-\varpi \big)\\\nonumber
&+\varpi\big)+B \big(\varpi  q^2+r^4 (2-24 \varpi  \mathfrak{B_c} )\big)\big\}\bigg]^{-1}\bigg[4 \big\{\varpi  A^2 \mathfrak{B_c}  r^6
+A \big(2 \varpi  \mathfrak{B_c}  B r^6-\varpi  q^2\\\nonumber
&+r^4 (7 \varpi  \mathfrak{B_c} -2)\big)-B\big(3 \varpi  \mathfrak{B_c}  B r^6+\varpi  q^2+r^4 (2-\varpi  \mathfrak{B_c} )\big)\big\}\bigg].
\end{align}
The term $\big|v_{s\bot}^2-v_{sr}^2\big|$ in modified theory takes the form
\begin{align}\nonumber
\big|v_{s\bot}^2-v_{sr}^2\big|&=\frac{1}{12}\bigg|\bigg[\varpi  A^3 r^4 \big(4 \mathfrak{B_c}  r^6 e^{A r^2}
+r^4 \big(2-4 \varpi  \mathfrak{B_c}  \big(e^{A r^2}-2\big)\big)+3\varpi  q^2\big)\\\nonumber
&+A^2 r^2 \big\{4 e^{A r^2} \big(2 \varpi  \mathfrak{B_c}  B r^8-2 r^6 \big(-3 \varpi  \mathfrak{B_c}+\varpi ^2 \mathfrak{B_c}  B+1\big)
+\varpi ^2 q^2\\\nonumber
&-\varpi  q^2 r^2+\varpi  r^4 (2-5 \varpi  \mathfrak{B_c} )\big)+\varpi\big(B \big(33 \varpi  q^2 r^2+r^6 (22
-64 \varpi  \mathfrak{B_c} )\big)\\\nonumber
&+30 \varpi  q^2-8 \varpi  \mathfrak{B_c}r^4\big)\big\}-A \big\{\varpi  B^2 r^4 \big(12 \mathfrak{B_c}  r^6 e^{A r^2}
-2 r^4 \big(2 \varpi  \mathfrak{B_c}\big(3e^{A r^2}\\\nonumber
&-20\big)+7\big)-21 \varpi  q^2\big)+4 B r^2 \big(e^{A r^2} \big(\varpi  q^2r^2-\varpi ^2 q^2+r^6 (\varpi  \mathfrak{B_c} +2)\\\nonumber
&-\varpi  r^4 (3 \varpi  \mathfrak{B_c} +2)\big)+4 \varpi ^2 \big(q^2+\mathfrak{B_c}r^4\big)\big)
-4 \varpi  \big(e^{A r^2} \big(q^2 \big(\varpi -2 r^2\big)\\\nonumber
&+r^4 (7 \varpi  \mathfrak{B_c}-2)\big)-\varpi  q^2+r^4 (2-7 \varpi  \mathfrak{B_c} )\big)\big\}
+\varpi  B \big\{4 e^{A r^2} \big(r^4\big(\varpi  \mathfrak{B_c}\\\nonumber
&+3 \mathfrak{B_c}  B r^2 \big(r^2-2 \varpi \big)-2\big)+q^2 \big(\varpi -2r^2\big)\big)
-r^4 \big(4 \varpi  \mathfrak{B_c}+9 \varpi  B^2 q^2\\\nonumber
&-8\big)+6 B^2 r^8 (12 \varpi  \mathfrak{B_c} -1)-46\varpi  B q^2 r^2+24 \varpi  \mathfrak{B_c}  B r^6-4 \varpi  q^2\big\}\bigg]^{-1}\\\nonumber
&\times\bigg[-4+\bigg\{r^3 e^{-A r^2} \big(\varpi  \big(-A^2 r^4-5 A r^2 \big(2 B r^2+3\big)+23 B r^2+4\\\nonumber
&+3 B^2 r^4\big)+4 e^{Ar^2} \big(r^2-\varpi \big)\big)^2\bigg(-\frac{2q^2}{r^3}\big(-\varpi A B+2 A e^{A r^2}+\varpi B^2\big)\\\nonumber
&+4Br\big(\varpi e^{-A r^2} \big(-A \big(Br^2+2\big)+\frac{1-e^{A r^2}}{r^2}+B^2 r^2+3 B\big)+1\big)^{-1}\\\nonumber
&\times(B-A)-\frac{4q^2}{r^5}\big(\varpi\big(A B r^2+A-B \big(B r^2+2\big)\big)-2 e^{A r^2}\big)+\big(4 \varpi  r\\\nonumber
&\times e^{A r^2} \big(A B r^2+A-B \big(B r^2+2\big)\big) \big(A^2 r^4 \big(B r^2+2\big)-A\big(B^2 r^6+r^2\\\nonumber
&+4 B r^4\big)+e^{A r^2}+B^2 r^4-1\big)\big)\big(\varpi  \big(-A r^2 \big(B r^2+2\big)+B^2 r^4+3 Br^2\\\nonumber
&+1\big)+e^{A r^2} \big(r^2-\varpi\big)\big)^{-2}+\big(32 \varpi  r^3 e^{A r^2} \big(\frac{1}{4 r^4}\big((A+B) \big(3 A^2 r^2+A\\\nonumber
&\times\big(B r^2-2\big)+2 B \big(Br^2-2\big)\big) \big(\varpi  q^2+2 r^4\big)\big)+\mathfrak{B_c}  e^{A r^2} \big(-3 A^2 r^2+A\\\nonumber
&\times\big(Br^2+2\big)-2 B\big)+\frac{1}{2 r^2}\big(\varpi  \mathfrak{B_c}  \big(A^3 r^4 \big(11 B r^2+12\big)-A^2 r^2 \big(7 B^2 r^4\\\nonumber
&-6 e^{Ar^2}+52 B r^2+14\big)+B \big(4 e^{A r^2}+3 B^3 r^6-10 B^2 r^4-21 B r^2-4\big)\\\nonumber
&+A \big(-2 e^{Ar^2} \big(B r^2+2\big)+B^3 r^6+18 B^2 r^4+55 B r^2+4\big)\big)\big)\big)\big)\big(\big(\varpi\\\nonumber
&\times\big(-A r^2 \big(B r^2+2\big)+B^2 r^4+3 B r^2+1\big)+e^{A r^2} \big(r^2-\varpi\big)\big) \big(\varpi  \big(-A^2\\\nonumber
&\times r^4-5 A r^2 \big(2 B r^2+3\big)+3 B^2 r^4+23 B r^2+4\big)+4 e^{Ar^2} \big(r^2-\varpi \big)\big) \big)^{-1}\\\nonumber
&+\big(16 \varpi  A r^5 e^{A r^2} \big(\frac{1}{4 r^4}\big((A+B) \big(3 A^2 r^2+A \big(B r^2-2\big)+2 B \big(Br^2-2\big)\big)\\\nonumber
&\times\big(\varpi  q^2+2 r^4\big)\big)+\mathfrak{B_c}  e^{A r^2} \big(-3 A^2 r^2+A \big(Br^2+2\big)-2 B\big)
+\frac{1}{2 r^2}\big(\varpi\\\nonumber
&\times\mathfrak{B_c}  \big(A^3 r^4 \big(11 B r^2+12\big)-A^2 r^2 \big(-6 e^{Ar^2}+7 B^2 r^4+52 B r^2+14\big)\\\nonumber
&+B \big(4 e^{A r^2}+3 B^3 r^6-10 B^2 r^4-21 B r^2-4\big)+A \big(-2 e^{Ar^2} \big(B r^2+2\big)\\\nonumber
&+B^3 r^6+18 B^2 r^4+55 B r^2+4\big)\big)\big)\big) \big)\big(\big(\varpi  \big(-A r^2 \big(B r^2+2\big)+B^2 r^4\\\nonumber
&+3 Br^2+1\big)+e^{A r^2} \big(r^2-\varpi \big)\big) \big(\varpi
\big(-A^2 r^4-5 A r^2 \big(2 B r^2+3\big)+4\\\nonumber
&+3 B^2 r^4+23 Br^2\big)+4 e^{A r^2} \big(r^2-\varpi \big)\big) \big)^{-1}-\big(16 \varpi  r^5 e^{A r^2} \big(\varpi
\big(B \big(2 B r^2\\\nonumber
&+3\big)-2 A\big(B r^2+1\big)\big)+e^{A r^2} \big(A \big(r^2-\varpi\big)+1\big)\big) \big(\frac{1}{4 r^4}\big((A+B) \big(3 A^2 r^2\\\nonumber
&+A \big(Br^2-2\big)+2 B \big(B r^2-2\big)\big) \big(\varpi  q^2+2r^4\big)\big)+\mathfrak{B_c}  e^{A r^2} \big(-3 A^2 r^2\\\nonumber
&+A \big(Br^2+2\big)-2 B\big)+\frac{1}{2 r^2}\big(\varpi  \mathfrak{B_c}  \big(A^3 r^4 \big(11 Br^2+12\big)-A^2 r^2 \big(14\\\nonumber
&-6 e^{A r^2}+7 B^2 r^4+52 B r^2\big)+B\big(4 e^{A r^2}+3 B^3 r^6-10 B^2 r^4-21 B r^2\\\nonumber
&-4\big)+A \big(-2e^{A r^2} \big(B r^2+2\big)+B^3 r^6+18 B^2 r^4+55 Br^2+4\big)\big)\big)\big) \big)\\\nonumber
&\times\big(\big(\varpi  \big(-A r^2 \big(B r^2+2\big)+B^2 r^4+3 B r^2+1\big)+e^{A r^2} \big(r^2-\varpi\big)\big)^2\big(\varpi\\\nonumber
&\times\big(-A^2 r^4-5 A r^2 \big(2 B r^2+3\big)+3 B^2 r^4+23 B r^2+4\big)+4\big(r^2-\varpi \big)\\\nonumber
&\times e^{A r^2} \big) \big)^{-1}-\big(16 \varpi  r^5 e^{A r^2} \big(\varpi  \big(-2 A^2 r^2-5 A \big(4 B r^2+3\big)+B \big(6 Br^2\\\nonumber
&+23\big)\big)+4 e^{A r^2} \big(A \big(r^2-\varpi \big)+1\big)\big) \big(\frac{1}{4 r^4}\big((A+B) \big(3A^2 r^2+A \big(B r^2\\\nonumber
&-2\big)+2 B \big(B r^2-2\big)\big) \big(\varpi  q^2+2 r^4\big)\big)+\beta e^{A r^2} \big(-3 A^2 r^2+A \big(B r^2+2\big)\\\nonumber
&-2 B\big)+\frac{1}{2 r^2}\big(\varpi  \beta  \big(A^3 r^4 \big(11 Br^2+12\big)-A^2 r^2 \big(-6 e^{A r^2}+7 B^2 r^4\\\nonumber
&+52 B r^2+14\big)+B \big(4 e^{A r^2}+3 B^3 r^6-10 B^2r^4-21 B r^2-4\big)+A\\\nonumber
&\times\big(-2 e^{A r^2} \big(B r^2+2\big)+B^3 r^6+18 B^2 r^4+55 Br^2+4\big)\big)\big)\big)\big)\big(\big(\varpi\\\nonumber
&\times\big(-A r^2 \big(B r^2+2\big)+B^2 r^4+3 B r^2+1\big)+e^{A r^2} \big(r^2-\varpi\big)\big) \big(\varpi  \big(-A^2\\\nonumber
&\times r^4-5 A r^2 \big(2 B r^2+3\big)+3 B^2 r^4+23 B r^2+4\big)+4e^{A r^2} \big(r^2-\varpi \big)\big)^2 \big)^{-1}\\\nonumber
&-Ar\bigg(\big(4(-A(1+Br^2)+B(2+Br^2))\big)\big(\varpi e^{-A r^2} \big(-A \big(B r^2+2\big)\\\nonumber
&+\frac{1-e^{A r^2}}{r^2}+B^2 r^2+3 B\big)+1\big)^{-1}+\frac{2q^2}{r^4}\big(\varpi\big(A B r^2-B \big(B r^2+2\big)\\\nonumber
&+A\big)-2 e^{A r^2}\big)+\big(16 \varpi  r^4 e^{A r^2} \big(\frac{1}{4 r^4}\big((A+B) \big(3 A^2 r^2+A \big(B r^2-2\big)\\\nonumber
&+2 B \big(Br^2-2\big)\big) \big(\varpi  q^2+2 r^4\big)\big)+\mathfrak{B_c}  e^{A r^2} \big(-3 A^2 r^2+A \big(Br^2+2\big)\\\nonumber
&-2 B\big)+\frac{1}{2 r^2}\big(\varpi  \mathfrak{B_c}  \big(A^3 r^4 \big(11 B r^2+12\big)-A^2 r^2 \big(-6 e^{Ar^2}+7 B^2 r^4\\\nonumber
&+52 B r^2+14\big)+B \big(4 e^{A r^2}+3 B^3 r^6-10 B^2 r^4-21 B r^2-4\big)+A\\\nonumber
&\times\big(-2 e^{Ar^2} \big(B r^2+2\big)+B^3 r^6+18 B^2 r^4+55 B r^2+4\big)\big)\big)\big)\big)\big(\big(\varpi\\\nonumber
&\times\big(-A r^2 \big(B r^2+2\big)+B^2 r^4+3 B r^2+1\big)+e^{A r^2} \big(r^2-\varpi\big)\big) \big(\varpi\\\nonumber
&\times\big(-A^2 r^4-5 A r^2 \big(2 B r^2+3\big)+3 B^2 r^4+23 B r^2+4\big)+4\big(r^2-\varpi \big)\\\nonumber
&\times e^{A r^2}\big)\big)^{-1}\bigg)+\big(4 \varpi  e^{A r^2} \big(A^3 r^2 \big(-4 \mathfrak{B_c}  r^6 \big(3 e^{A r^2}
-11 \varpi  B\big)+6 r^4 \big(2\varpi\\\nonumber
&\times\mathfrak{B_c}  \big(e^{A r^2}+2\big)+1\big)-3 \varpi  q^2\big)-4 A^2 \big(B \big(\mathfrak{B_c}  r^8\big(-e^{A r^2}\big)
+r^6 \big(\varpi\mathfrak{B_c}\\\nonumber
&\times\big(e^{A r^2}+26\big)-2\big)+\varpi  q^2r^2\big)+\mathfrak{B_c}  r^4 e^{A r^2} \big(2 \varpi +r^2\big)
+7 \varpi  \mathfrak{B_c}  B^2 r^8\\\nonumber
&-\varpi  q^2\big)+2 B\big(-4 \varpi  \mathfrak{B_c}  r^2 \big(e^{A r^2}-1\big)+6 \varpi  \mathfrak{B_c}  B^3 r^8
-B^2 \big(\varpi  q^2 r^2\\\nonumber
&+2r^6 (5 \varpi  \mathfrak{B_c} -1)\big)+4 \varpi  B q^2\big)+A \big(4 B \big(3 \varpi  q^2
-\mathfrak{B_c}  r^4 e^{A r^2}\big(r^2-2 \varpi \big)\big)\\\nonumber
&+8 \varpi  \mathfrak{B_c}  r^2 \big(e^{A r^2}-1\big)+B^2\big(6 r^6 (6 \varpi  \mathfrak{B_c} +1)-3 \varpi  q^2 r^2\big)
+4 \varpi  \mathfrak{B_c}\\\nonumber
&\times B^3 r^8\big)\big)\big)\big(r \big(\varpi  \big(A^2 r^4+5 A r^2 \big(2 B r^2+3\big)-3 B^2 r^4-23 B r^2-4\big)\\\nonumber
&-4 e^{A r^2}\big(r^2-\varpi \big)\big) \big(\varpi  \big(A r^2 \big(B r^2+2\big)-B^2 r^4-3 B r^2-1\big)+e^{Ar^2}\\\nonumber
&\times\big(\varpi -r^2\big)\big) \big)^{-1}  \bigg)\bigg\}\bigg]\bigg|.
\end{align}

\end{document}